%% file: paper-1.tex
\documentclass[11pt,a4paper]{article}

\usepackage{fontspec}
\defaultfontfeatures{Ligatures=TeX}
\usepackage[a4paper,left=22mm,right=22mm,top=30mm,bottom=30mm,headheight=40pt,headsep=20pt]{geometry}
\usepackage{microtype}
\usepackage{graphicx}
\usepackage{pdflscape}
\usepackage{booktabs}
\usepackage{longtable}
\usepackage{array}
\usepackage[table]{xcolor}
\usepackage{colortbl}
\usepackage{calc}
\usepackage{ragged2e}
\usepackage{enumitem}
\usepackage{fancyhdr}
\usepackage[explicit]{titlesec}
\usepackage{needspace}
\usepackage{caption}
\usepackage{float}
\usepackage{wrapfig}
\usepackage{afterpage}
\usepackage{adjustbox}
\usepackage{tcolorbox}
\usepackage{tikz}
\usepackage{listings}
\usepackage{natbib}
\usepackage{xurl}
\usepackage{hyperref}

\makeatletter
\define@key{Gin}{alt}{}
\define@key{Gin}{inkscapelatex}{}
\makeatother

\definecolor{KinroGraphite}{HTML}{262624}
\definecolor{KinroAsh}{HTML}{46453F}
\definecolor{KinroRock}{HTML}{C9C7BC}
\definecolor{KinroStone}{HTML}{E7E6DD}
\definecolor{KinroPebble}{HTML}{F3F1EA}
\definecolor{KinroOffWhite}{HTML}{FEFEFB}
\definecolor{KinroHighVis}{HTML}{F2FF5A}
\definecolor{KinroCobalt}{HTML}{0022FF}
\definecolor{KinroPlotSMS}{HTML}{2563EB}
\definecolor{KinroPlotEmail}{HTML}{7C3AED}
\definecolor{KinroPlotCalls}{HTML}{D97706}
\definecolor{KinroPlotInk}{HTML}{17211C}
\definecolor{KinroPlotPath}{HTML}{CBD5CE}
\definecolor{KinroPlotGrid}{HTML}{E0E8E2}
\definecolor{KinroPlotLabel}{HTML}{647269}
\definecolor{KinroLossOne}{HTML}{0022FF}
\definecolor{KinroLossTwo}{HTML}{6D28D9}
\definecolor{KinroLossThree}{HTML}{0F766E}
\definecolor{KinroLossFour}{HTML}{D97706}
\definecolor{KinroLossFive}{HTML}{BE123C}
\definecolor{KinroLossSix}{HTML}{2563EB}
\definecolor{KinroLossSeven}{HTML}{7C3AED}
\definecolor{KinroLossEight}{HTML}{64748B}
\definecolor{KinroLossNine}{HTML}{C9C7BC}
\definecolor{KinroLossTen}{HTML}{334155}

\newcommand{\kinrolegenddot}[2]{%
  \tikz[baseline=-0.55ex]{\filldraw[fill=#1,draw=#1] (0,0) circle[radius=1.45pt];}%
  \hspace{0.25em}#2%
}
\newcommand{\kinrolegendinbound}{%
  \tikz[baseline=-0.55ex]{\filldraw[fill=white,draw=KinroPlotInk,line width=0.6pt] (0,0) circle[radius=1.45pt];}%
  \hspace{0.25em}inbound%
}
\newcommand{\kinrolegendbound}{%
  \tikz[baseline=-0.55ex]{\path[fill=KinroPlotInk] (0,2pt) -- (2pt,0) -- (0,-2pt) -- (-2pt,0) -- cycle;}%
  \hspace{0.25em}bound%
}
\newcommand{\kinrocommunicationlegend}{%
  \par\noindent\centering\sffamily\fontsize{7}{8}\selectfont\color{KinroPlotLabel}%
  \kinrolegenddot{KinroPlotSMS}{SMS}\hspace{1.35em}%
  \kinrolegenddot{KinroPlotEmail}{Email}\hspace{1.35em}%
  \kinrolegenddot{KinroPlotCalls}{Calls}\hspace{1.35em}%
  \kinrolegenddot{KinroPlotInk}{outbound}\hspace{1.35em}%
  \kinrolegendinbound\hspace{1.35em}%
  \kinrolegendbound\par%
}

\hypersetup{
  colorlinks=true,
  linkcolor=KinroCobalt,
  citecolor=KinroCobalt,
  urlcolor=KinroCobalt,
  pdftitle={An Insurance Broker for Every Small Business: The Economics of Exceptional Care at Scale},
  pdfauthor={Kinro Team}
}

\newcommand{\kinroprimarybullet}{\textcolor{KinroCobalt}{\raisebox{0.14ex}{\scriptsize\textbullet}}}
\newcommand{\kinrosecondarybullet}{\textcolor{KinroAsh}{\raisebox{0.35ex}{\scriptsize\textendash}}}
\newcommand{\kinroinlinebullet}{\kinroprimarybullet\hspace{0.4em}}
\setlist[itemize,1]{
  label=\kinroprimarybullet,
  leftmargin=1.45em,
  labelsep=0.55em,
  itemsep=0.16em,
  topsep=0.35em,
  parsep=0.04em,
  partopsep=0pt
}
\setlist[itemize,2]{
  label=\kinrosecondarybullet,
  leftmargin=1.35em,
  labelsep=0.5em,
  itemsep=0.1em,
  topsep=0.2em,
  parsep=0pt,
  partopsep=0pt
}
\setlist[enumerate]{leftmargin=1.6em,itemsep=0.08em,topsep=0.2em,parsep=0pt}

\newcommand{\kinrosection}[1]{\color{KinroGraphite}#1}
\pretocmd{\section}{\Needspace{5\baselineskip}}{}{}
\pretocmd{\subsection}{\Needspace{4\baselineskip}}{}{}
\pretocmd{\subsubsection}{\Needspace{3\baselineskip}}{}{}
\titleformat{\section}{\fontsize{13}{14}\selectfont\bfseries}{\thesection.}{0.5em}{\kinrosection{#1}}
\titlespacing*{\section}{0pt}{3ex plus 4pt minus 3pt}{5pt}
\titleformat{\subsection}{\normalsize\bfseries}{\thesubsection}{0.55em}{\kinrosection{#1}}
\titlespacing*{\subsection}{0pt}{2.5ex plus 3pt minus 2pt}{2pt}
\titleformat{\subsubsection}{\normalsize\bfseries}{\thesubsubsection}{0.55em}{\kinrosection{#1}}
\titlespacing*{\subsubsection}{0pt}{2ex plus 2.5pt minus 1.5pt}{2pt}

\renewcommand{\headrulewidth}{1pt}
\renewcommand{\footrulewidth}{1pt}
\makeatletter
\renewcommand{\headrule}{\color{KinroRock}\hrule\@height\headrulewidth\@width\headwidth\vskip-\headrulewidth}
\renewcommand{\footrule}{\color{KinroRock}\hrule\@width\headwidth\@height\footrulewidth\vskip 2.6pt}
\makeatother
\fancypagestyle{firstpage}{
  \fancyhf{}
  \fancyhead[L]{%
    \includegraphics[height=22pt]{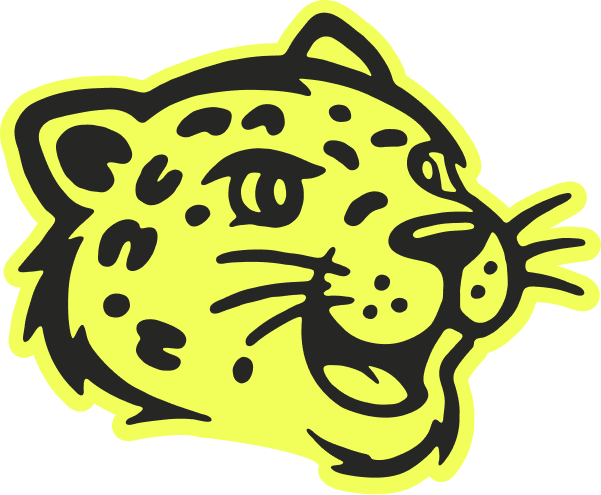}%
    \hspace{6pt}%
    \includegraphics[height=14pt]{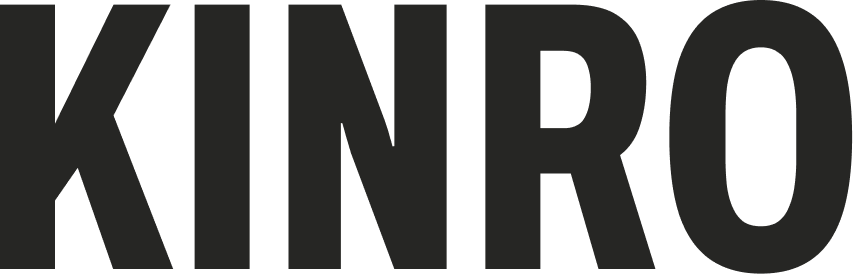}%
  }
  \fancyhead[R]{%
    \fontsize{8}{10}\selectfont\color{KinroGraphite}%
    \href{https://kinro.com}{kinro.com}%
  }
  \fancyfoot[L]{%
    \fontsize{7.5}{9.5}\selectfont\color{KinroGraphite}%
    \textsuperscript{*}See the \hyperref[contributions-and-acknowledgments]{Contributions and Acknowledgments}
    section for the full author list. Please send correspondence to
    \href{mailto:research@kinro.com}{research@kinro.com}.%
  }
  \renewcommand{\headrulewidth}{1pt}
  \renewcommand{\footrulewidth}{1pt}
}

\DeclareCaptionLabelSeparator{kinropipe}{\enspace|\enspace}
\newtcolorbox{kinroexample}{
  colback=KinroPebble,
  colframe=KinroRock,
  boxrule=0.6pt,
  arc=2mm,
  left=4mm,
  right=4mm,
  top=3mm,
  bottom=3mm,
  before skip=0.55em,
  after skip=0.55em
}

\providecommand{\tightlist}{\setlength{\itemsep}{0pt}\setlength{\parskip}{0pt}}

\renewcommand{\arraystretch}{1.2}
\arrayrulecolor{KinroRock}
\setcitestyle{numbers,square,comma}

\begin{document}

\thispagestyle{firstpage}
{\fontsize{18}{20}\selectfont\bfseries\color{KinroGraphite}\raggedright
An Insurance Broker for Every Small Business: The Economics of Exceptional
Care at Scale\par}
\vspace{0.9em}
{\noindent\fontsize{9}{13}\selectfont\bfseries\color{KinroGraphite}Kinro
Team\textsuperscript{*}\par}
\vspace{1.05em}

{\fontsize{10}{12}\selectfont\bfseries\linespread{1.2}\selectfont
\noindent Small-business owners need expert guidance on their own terms, across
schedules, languages, and channels, but low premiums make exceptional,
continuous human service uneconomic for much of the market. Combining public
evidence, Kinro operational data, and an illustrative five-year service model,
we show why traditional brokerage economics leave 35 million U.S. small businesses
underserved. An AI-native brokerage can change those economics by performing and
coordinating routine work continuously, while licensed professionals
govern consequential exceptions and the brokerage remains accountable.\par}
\vspace{1em}

{\fontsize{11}{12}\selectfont\itshape\color{KinroGraphite}\noindent
\par}

\setlength{\parskip}{0.22\baselineskip plus 0.04\baselineskip minus 0.02\baselineskip}

\section{Introduction}\label{introduction}

Small businesses are widely underinsured. Many lack a broker who can
understand their operations, search available markets, compare coverage and
price, complete placement, and provide continuing service. Their risks matter
no less than those of larger companies, but small commissions cannot fund the
same attention. Traditional brokerage economics therefore ration service by
account profitability even though the owner's need does not shrink with the
commission.

\begin{figure}[H]
\centering
\resizebox{\linewidth}{!}{\input{assets/current-customer-annual-commission-distribution.tex}}
\captionsetup{justification=justified,singlelinecheck=false}
\caption{\textbf{Kinro operational evidence.} Annual gross commission per customer for policies Kinro placed during the previous three months. AI lines show annualized direct-cost break-even for GPT-5.6 Luna (\$31) and Sol (\$41), including \$150 in non-compute acquisition spend per bound customer (marketing and lead generation), full-funnel AI usage, and five years of service. Human lines show the commission required for a 50\% contribution margin: \$444 for basic service and \$1,020 for high-touch service, using the same acquisition assumption and Table~\ref{tab:five-year-cost}'s service costs. The left panel expands the under-\$100 band.}
\label{fig:current-customer-commission-distribution}
\end{figure}
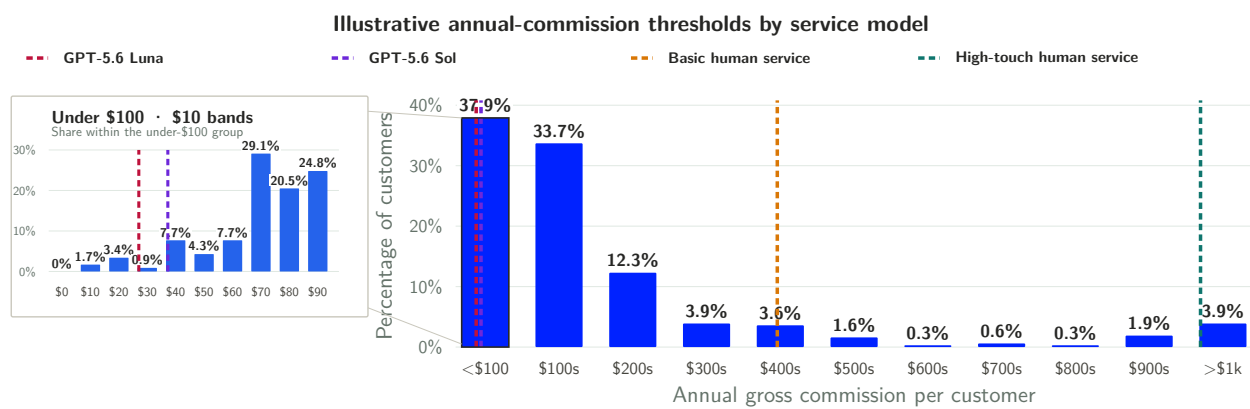

For three months, Kinro did not discriminate among accounts based on expected
premium and applied the same service standard to more than 3,000 insurance
buyers.
Figure~\ref{fig:current-customer-commission-distribution} shows the result.
Using the cost assumptions in
Table~\ref{tab:five-year-cost} and a 50\% contribution-margin target, a
basic-service broker would need \$444 in annual commission and would not have
served about 90\% of observed accounts for economic reasons; a high-touch
broker would need \$1,020 and would not have served at least 96\%.

\href{https://kinro.com}{Kinro} is a licensed insurance brokerage whose AI agents
coordinate customer communication, intake, quote workflows, follow-up, and
routine service (more on Kinro in Section~\ref{about-kinro}). Because
low-commission accounts can fund only a few human touchpoints, continuous
service ultimately requires bounded autonomy: the system completes
demonstrated work within explicit limits, while licensed professionals govern
those limits and resolve unfamiliar or high-risk decisions. This does not
authorize AI to independently perform every regulated act involved in selling,
underwriting, or binding insurance. Those activities remain within the
authority of the appropriate licensed brokerage, carrier, MGA, and
professionals. Kinro remains accountable and liable for its standard of care
and maintains E\&O insurance against that risk.

\textbf{Scope of this paper.} This paper examines why high-quality brokerage
service is economically unavailable to many small businesses and how a
different service model could change those economics. It does not establish
that AI can independently underwrite or price every risk, or that its effect on
claims and loss ratios is already known. Underwriting quality, premium
adequacy, and sustainable loss performance require separate evaluation with
carriers and MGAs using submission, underwriting, pricing, and mature claims
evidence.

This paper is intended for brokers, carriers, MGAs, wholesale markets,
regulators, and consumer advocates. We invite them to form an industry-wide
collective advancing safe, compliant AI-enabled brokerage that improves the
customer experience, extends service to businesses unable to afford adequate
help, and respects carrier appetite. Through bounded real-world pilots, the
collective can establish common benchmarks, measure customer and underwriting
outcomes, and determine when broader AI authority is warranted.
Section~\ref{conclusion-and-industry-agenda} presents our current view of how
each participant could contribute.

\section{Why small businesses are
underserved}\label{why-small-businesses-are-underserved}

\subsection{Owners have no insurance
department}\label{owners-have-no-insurance-department}

\begin{wrapfigure}[12]{r}{0.49\textwidth}
\vspace{-0.75\baselineskip}
\centering
\includegraphics[width=\linewidth]{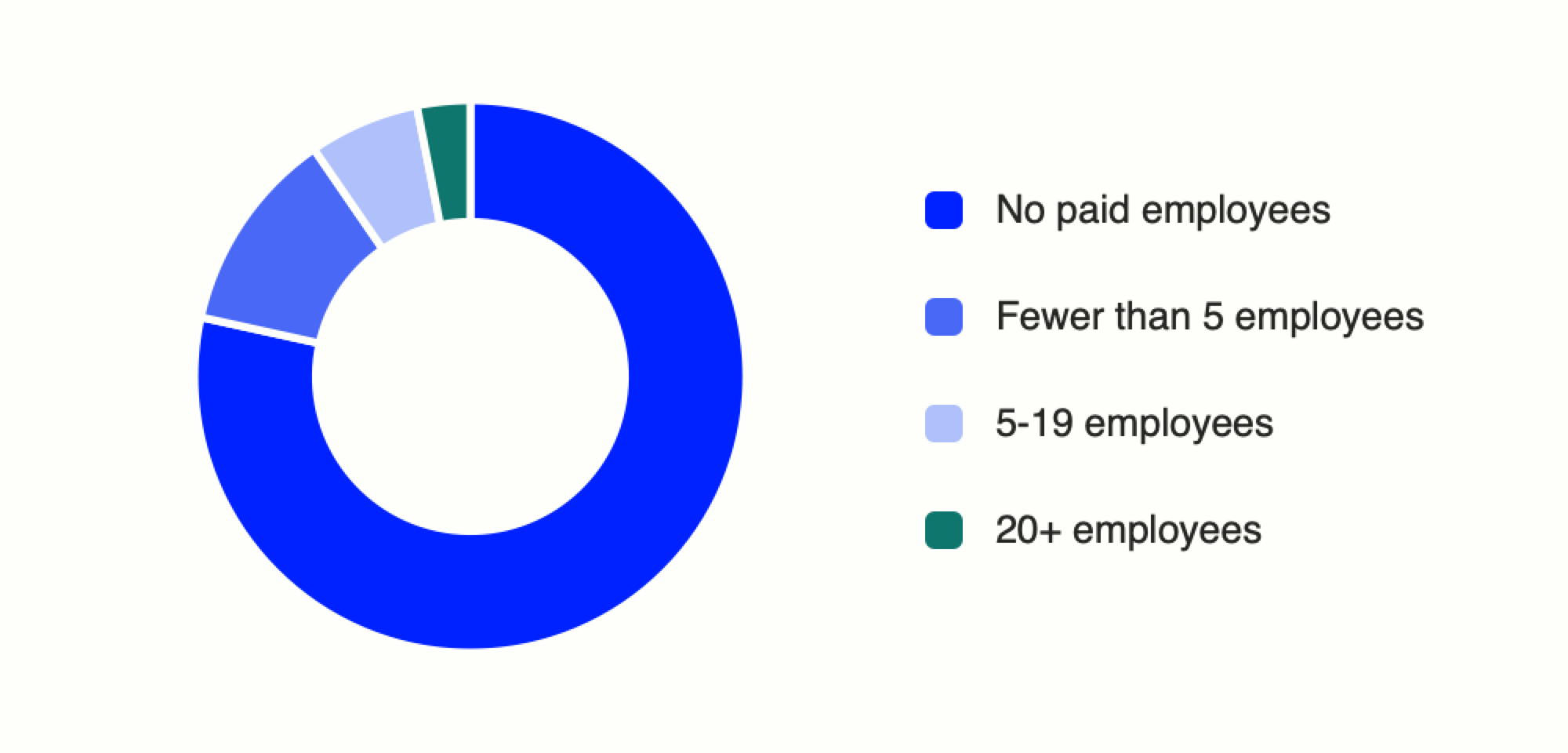}
\caption{\textbf{Public evidence.} 90.4\% of U.S. establishments had fewer
than five paid employees in 2023. That was 35.07 million of 38.79 million~\citep{census-cbp-2023,census-nonemployer-legal-form-2023}.}
\label{fig:us-business-establishments-by-employment-size}
\vspace{-0.5\baselineskip}
\end{wrapfigure}

Small-business owners have no department to hand problems to. They sell,
operate, and fix whatever breaks, while payroll, taxes, contracts, and
insurance keep demanding attention. Much of this work happens at night
or on weekends, at the expense of family, rest, and health.
Figure~\ref{fig:customer-messages} later shows this pattern in Kinro's
operations: 51.1\% of observed inbound customer messages arrived outside
weekday business hours.

This is the typical business, not an edge case. In 2023, establishments with
no paid employees made up 78.4\% of all U.S. business establishments, and
90.4\% had either no paid employees or fewer than five. These businesses also
form an operating backbone beneath larger enterprises: owners supply goods and
specialized services as vendors, contractors, and subcontractors. The federal
contracting system makes this dependence explicit by requiring certain large
prime contractors to create opportunities for small-business subcontractors~\citep{sba-prime-subcontracting}.

\subsection{Small-business risk is highly
heterogeneous}\label{small-business-risk-is-highly-heterogeneous}

Small businesses are not a single risk class. The 2022 North American Industry
Classification System divides the U.S. economy into 20 sectors and 1,012
national industries based on differences in how establishments produce goods
or provide services~\citep{census-naics-2022-manual}. Employee count and
revenue therefore do not, by themselves, describe what a business does or the
risks its operations create. Businesses of similar size may have fundamentally
different property, liability, vehicle, professional, cyber, or workplace
exposures.

Insurance work must follow those operational differences. Discovery questions,
classifications, coverage needs, underwriting information, and available
markets can all change with the work performed, where it is performed, the
customers served, the property or equipment used, and the obligations assumed.
The U.S. Small Business Administration similarly advises owners to assess their
particular risks and identifies different forms of coverage for manufacturers,
service businesses, property-intensive operations, and home-based
businesses~\citep{sba-business-insurance}.

Specialization can make one category more repeatable, but it does not solve the
market-wide problem. Small-business demand is fragmented across industries,
geographies, and moments of need. A broker may not encounter enough similar
accounts, at a sufficiently low acquisition cost, to sustain a dedicated
workflow for every niche. Broadening the set of businesses served restores
volume but reintroduces variation. The least common businesses are therefore
especially likely to fall between generalized distribution and specialized
expertise.

\subsection{Small businesses form a large insurance
market}\label{small-businesses-form-a-large-insurance-market}

Their aggregate insurance market is substantial even though each account is
small. Deloitte's 2024 analysis estimates \$32.8 billion in annual
premium from 27.1 million nonemployer businesses and another \$7.3
billion from 3.8 million businesses with one to four employees. Together,
businesses with fewer than five paid employees represent an estimated
\$40.1 billion in annual premium across 30.9 million accounts. The
four employee-size segments in Deloitte's figure total \$74.1 billion,
effectively the report's approximately \$74 billion standard, non-specialty
portion of a \$113 billion under-50-employee commercial market. On that basis,
under-five businesses generate about 54\% of standard small-commercial
premium while representing 94\% of the businesses~\citep{deloitte-small-commercial-2024}.

The estimate may understate the market at today's business count. Deloitte's
analysis uses 30.9 million businesses with fewer than five employees, compared
with 35.07 million in the 2023 Census populations shown in Figure
\ref{fig:us-business-establishments-by-employment-size}. Scaling Deloitte's two
subsegments to those newer business counts would imply approximately
\$45.7 billion in annual premium. This is a count-updated sensitivity,
not observed written premium: insurer filings do not segment premiums by the
insured's employee count, and Deloitte's estimates combine SNL Financial,
Census, and internal data~\citep{deloitte-small-commercial-2024,census-cbp-2023,census-nonemployer-legal-form-2023}.

\subsection{Insurance becomes urgent before it becomes
understood}\label{insurance-becomes-urgent-before-it-becomes-understood}

Insurance is a consequential responsibility that few owners address on
their own timetable. It often becomes urgent after a loss or close call,
or when a customer, lender, landlord, or platform demands proof of
coverage. The owner must then decide quickly, often without the
knowledge needed to judge the options.

Insurance protects the business an owner has built by transferring
losses it may not be able to absorb. Without suitable coverage, a single
event can end the business. Property interruptions~\citep{katrina-recovery,christchurch-interruption,harvey-distress},
liability claims
~\citep{sba-litigation,malpractice-limits,malpractice-exit,rand-liability},
workplace injuries
~\citep{injury-survival,workplace-inspections,california-workers-comp,virginia-workers-comp,massachusetts-workers-comp},
and cyberattacks
~\citep{mastercard-cyber,gao-cyber-insurance,ftc-cyber-insurance} can
consume working capital, halt operations, or create direct legal
obligations.

Hiscox\textquotesingle s 2025 survey~\citep{hiscox-underinsurance}
classified 77\% of the participating US small businesses as underinsured
and found widespread misunderstanding of what common policies cover. The
broader protection gap takes two forms: no insurance at all, or
insurance whose limits, exclusions, deductibles, or conditions do not
match the losses the business could face.

\Needspace{0.48\textheight}
\subsection{Demand is growing as expertise
declines}\label{demand-is-growing-as-expertise-declines}

Figure \ref{fig:us-business-establishments-by-employment-size} shows that the
operating market is concentrated at the smallest end. Census application data
also point to a growing pipeline of prospective businesses.

\begin{wrapfigure}[15]{r}{0.49\textwidth}
\vspace{-0.75\baselineskip}
\centering
\includegraphics[width=\linewidth]{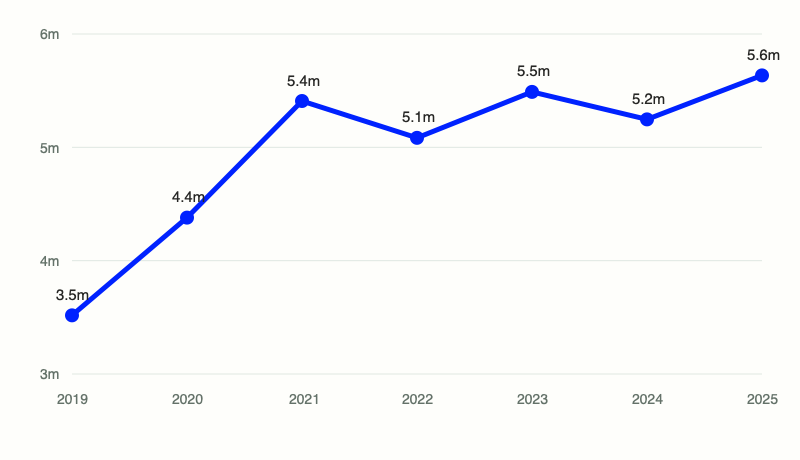}
\caption{\textbf{Public evidence.} Annual U.S. business applications, 2019--2025. Applications
rose 60.2\%, from 3.52 million to 5.63 million. Census Business Formation
Statistics count EIN applications, not confirmed businesses~\citep{census-business-formation-statistics,census-business-formation-faq}.}
\label{fig:us-business-applications-and-formations}
\vspace{-0.5\baselineskip}
\end{wrapfigure}

AI is likely to make some large organizations leaner while enabling smaller
teams to perform work that once required more people~\citep{nber-ai-productivity-workforce-2026}. This shift could make more small
teams economically viable, increasing the number of owners who must make
insurance decisions.

Retirements also put hard-won expertise at risk. The Institutes reports that
half of the current insurance workforce will retire by 2035, leaving more than
400,000 positions to replace. In its survey of risk-management and insurance
professionals, 73\% of respondents identified lost institutional knowledge as
the retirement wave's most significant effect. Expanding service therefore
requires not only reaching more businesses, but preserving and transmitting
the judgment of experienced professionals as they leave the workforce~\citep{institutes-retirement}.

\subsection{Today\textquotesingle s buying paths leave
gaps}\label{todays-buying-paths-leave-gaps}

\textbf{On their own.} Owners may complete unfamiliar applications repeatedly,
learning only at the end that a carrier will not insure their business. A
single carrier's answer does not establish the best available combination of
price and coverage, yet owners are expected to identify appropriate markets
and compare premiums, limits, exclusions, conditions, and product fit without
the necessary expertise.

\textbf{Through an agent or broker.} Some owners find a broker who truly
helps; others reach a seller representing one carrier, a broker with little
time for a small account, or no advisor at all. Lead-generation systems can
make the experience worse by distributing the same owner's information to
competing agencies without transferring the context already collected. The
owner receives repeated calls, re-explains the business, resubmits documents,
and restarts the application. What appears efficient to the distribution
system becomes fragmented and exhausting for the customer.

The experience becomes even worse when the business has a hard-to-place risk.

\textbf{With a hard-to-place risk.} Standard carriers may decline unusual
operations, while E\&S placement often requires wholesale access, manual
submissions, and specialist time that small premiums do not support. Figure
\ref{fig:partner-lost-reasons} shows that carrier appetite, coverage, and
product fit accounted for 27.0\% of Kinro's lost leads with a specific
documented outcome. This is distinct from price competitiveness: the problem
was whether an accessible market would write the risk or offer suitable terms.
For a broker already expecting little revenue from the account, that added
market search, wholesale coordination, and specialist work makes the risk even
less economic to pursue, although the owner's need for help is greater.
These results document gaps in Kinro's available placement routes, not proof
that no insurer would write the risks.

\Needspace{22\baselineskip}
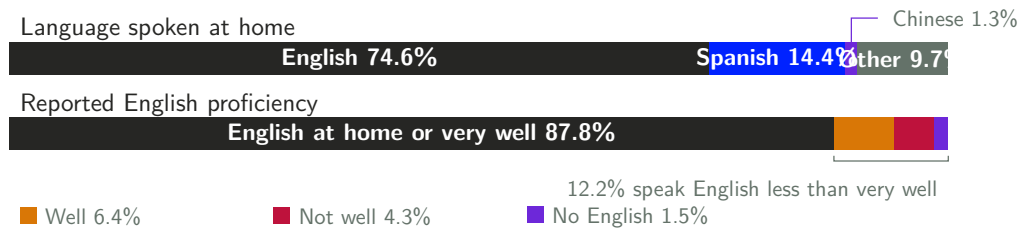
\begin{figure}[H]
\centering
\resizebox{0.82\linewidth}{!}{\input{assets/self-employed-language-profile.tex}}
\caption{\textbf{Public evidence. Language at home and English proficiency among self-employed
Americans, 2022.} 25.4\% spoke a language other than English at home; Spanish
accounted for 14.4\% and Chinese for 1.3\%. 12.2\% spoke English less than very
well, including 1.5\% who spoke no English~\citep{sba-language-business-ownership-2024}.}
\label{fig:self-employed-language}
\end{figure}

\textbf{In a language they trust.} Figure~\ref{fig:self-employed-language}
shows that this need is not marginal: one in four self-employed Americans
speaks a language other than English at home, and one in eight speaks English
less than very well. Owners need a broker who can navigate English-language
carrier systems while explaining consequential decisions in a language they
trust. That support matters even more during servicing and claims, when
carrier systems and internal teams may not support the customer's preferred
language.

\subsection{Service must follow the owner's operating rhythm and
language}\label{customer-behavior-requires-continuous-service}

Continuous service must adapt to each owner\textquotesingle s schedule, pace,
and preferred language.

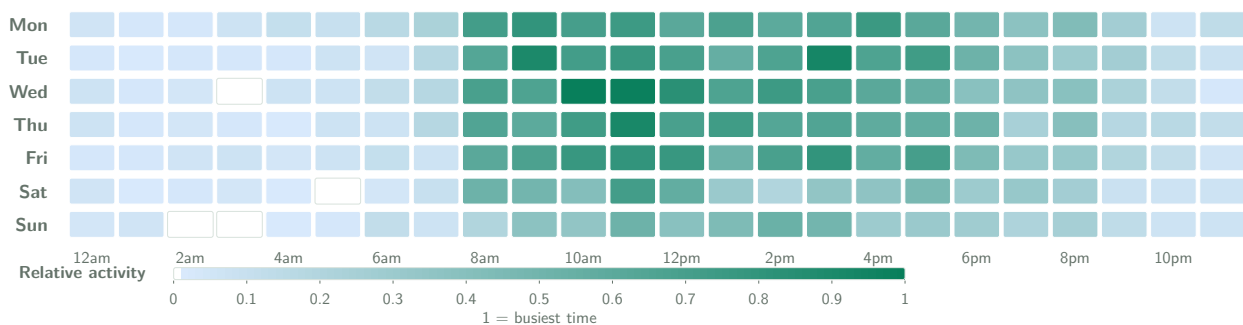
\begin{figure}[!b]
\centering
\input{assets/inbound-leads-by-local-week.tex}
\caption{\textbf{Kinro operational evidence.} Inbound customer messages arrive around the clock. In this snapshot, 51.1\% of inbound customer messages occurred outside Monday--Friday, 9am--5pm in the lead's local timezone. These messages follow outreach sent during permitted contact hours. Each cell counts distinct leads by lead-local weekday and hour; test and simulated activity are excluded.}
\label{fig:customer-messages}
\end{figure}

Customers from Hispanic and Chinese communities have asked Kinro to serve them
in Spanish and Chinese, and Kinro has sold policies in both languages.
These customers are not asking merely for forms to be translated. They want
to ask questions, understand tradeoffs, and complete the insurance process
in a language they trust. A broker must also adapt to the operating rhythm of
the owner's business. An owner serving customers, supervising a jobsite, or
moving between appointments may only be able to address insurance before the
day starts, between jobs, or after closing.

Figure~\ref{fig:customer-messages} shows that more than half of observed
inbound messages arrived outside weekday business hours. Availability at those
times is therefore not a convenience; it is part of serving owners on the
schedules their businesses allow. The chart aggregates businesses and does not
estimate industry-specific schedules, but it demonstrates why brokerage
service cannot be designed only around the broker's office hours.

\Needspace{4\baselineskip}
Channel preference also varies: some owners prefer SMS, some prefer email, and
some prefer phone calls. A broker must meet each customer in the channel they
prefer and preserve context when the conversation moves between channels.

\begin{figure}[H]
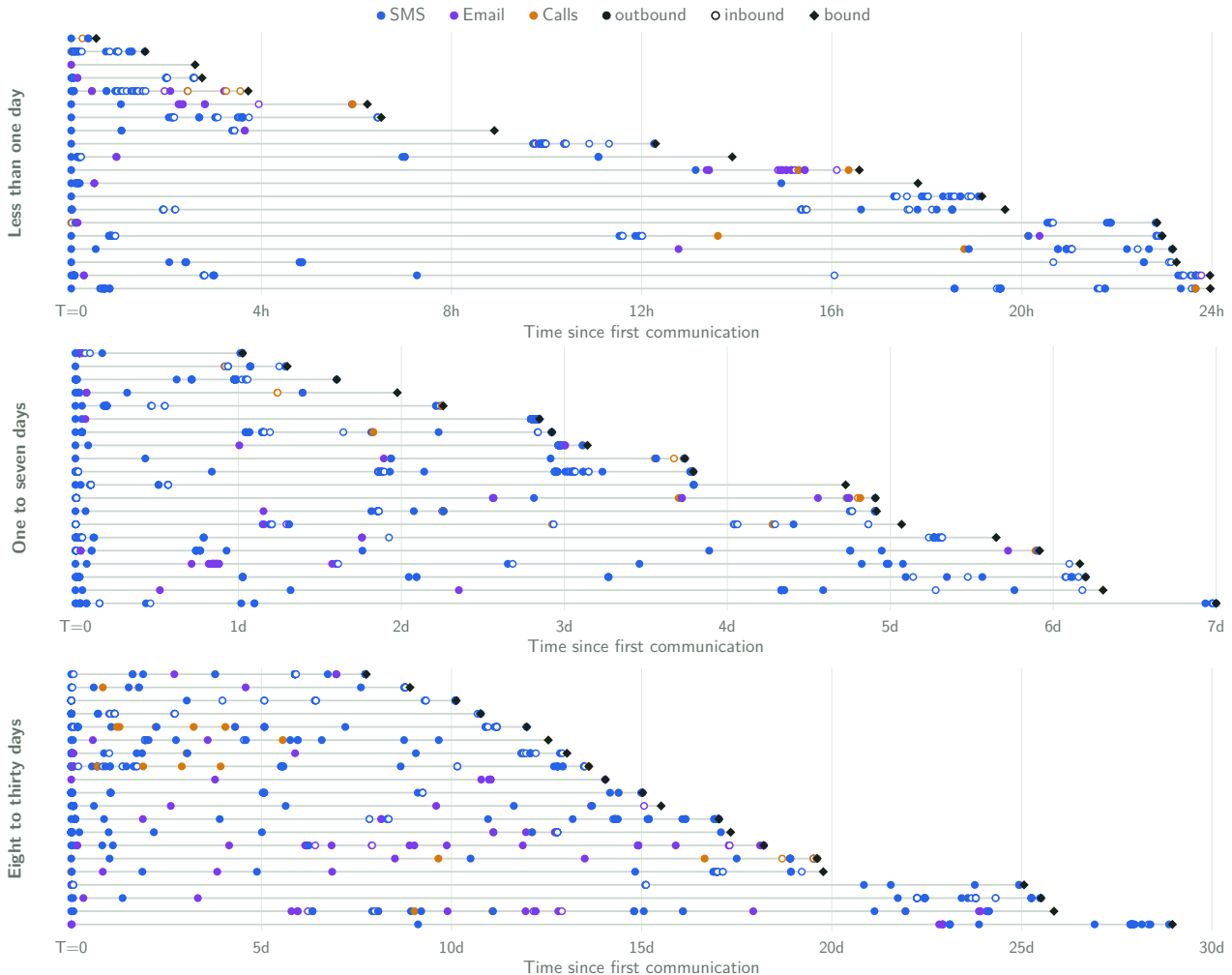

\centering
\kinrocommunicationlegend
\vspace{0.45em}
\input{assets/communications-to-close-same-day.tex}
\vspace{0.35em}
\input{assets/communications-to-close-one-to-seven-days.tex}
\vspace{0.35em}
\input{assets/communications-to-close-eight-to-thirty-days.tex}
\caption{\textbf{Kinro operational evidence.} Paths from first contact to first policy bind for a sample of 60 customers on different timelines. The three panels show customers who bound in less than one day, one to seven days, and eight to thirty days. Each row represents one customer, and each panel uses its own time scale.}
\label{fig:communication-paths}
\end{figure}

\textbf{Insurance work also moves at different tempos.}
Figure~\ref{fig:communication-paths} shows a sample of 60 customers divided
into three time-to-bind batches: less than one day, one to seven days, and
eight to thirty days.

The underlying business need can be immediate, such as satisfying a contract
or certificate deadline, or unfold as an owner gathers payroll, vehicle,
location, and prior coverage information around daily operations. Each panel
uses its own time scale. These durations begin with the first recorded SMS,
email, or answered call; they do not represent continuous time spent buying
insurance or isolate the effect of business type.

Figure~\ref{fig:communication-paths} makes a second point: outbound work often
continues after the customer has replied. Persistence can reassure owners that
the broker is still working for them and keep insurance from disappearing
behind the many other demands on their day. Because the figure shows a sample
of customers who ultimately bound, it does not prove that follow-up caused the
sale or that any one industry follows a particular timeline. It does show why
the service must preserve context, resume across interruptions, and maintain
reliable outbound follow-up at the pace each owner can sustain.

\subsection{Small-account economics limit
service}\label{small-account-economics-limit-service}

Small-business policies generate small commissions despite requiring sales,
advice, placement, and service. A \$600 policy at a 15\% commission yields
\$90 a year, yet still requires market search, certificates, renewals, and
answers.
The heterogeneity described in
Section~\ref{small-business-risk-is-highly-heterogeneous} limits the most
obvious human response: specialization. A niche broker can standardize
discovery and placement for one class of business, but must still acquire a
sufficient and predictable volume of those accounts. Expanding into adjacent
classes increases volume but reintroduces the variation in questions,
exposures, carrier appetite, and placement routes that specialization was meant
to remove.
Table~\ref{tab:five-year-cost} compares a high-touch broker, a basic-service
broker, and AI-enabled service priced with GPT-5.6 Sol and Luna across a
five-year relationship.
Section~\ref{why-ai-changes-small-account-economics}
explains the assumptions.

\begin{table}[H]
\centering
\footnotesize
\rowcolors{2}{white}{KinroPebble}
\begin{tabular}{@{}>{\raggedright\arraybackslash}p{0.28\linewidth} >{\raggedleft\arraybackslash}p{0.17\linewidth} >{\raggedleft\arraybackslash}p{0.17\linewidth} >{\raggedleft\arraybackslash}p{0.12\linewidth} >{\raggedleft\arraybackslash}p{0.12\linewidth}@{}}
\toprule
Customer work over five years & High-touch broker & Basic-service broker & GPT-5.6 Sol & GPT-5.6 Luna \\
\midrule
Sales and initial placement & 10h (\$400) & 8h (\$320) & \$12.00 & \$0.64 \\
Four annual renewals & 8h (\$320) & 4h (\$160) & \$7.20 & \$0.38 \\
Quarterly proactive market checks & 20h (\$800) & 0h (\$0) & \$28.80 & \$1.54 \\
Proactive customer check-ins & 10h (\$400) & 0h (\$0) & \$2.40 & \$0.13 \\
Certificates, routine service, and one claim & 12h (\$480) & 12h (\$480) & \$2.40 & \$0.13 \\
\textbf{Five years for one customer} & \textcolor{KinroCobalt}{\textbf{60h (\$2,400)}} & \textcolor{KinroCobalt}{\textbf{24h (\$960)}} & \textcolor{KinroCobalt}{\textbf{\$52.80}} & \textcolor{KinroCobalt}{\textbf{\$2.82}} \\
\bottomrule
\end{tabular}
\caption{\textbf{Illustrative model.} Direct cost of acquiring and serving one customer for five policy years. The sales and initial-placement row allocates full-funnel work across bound customers; Appendix A details the assumptions.}
\label{tab:five-year-cost}
\end{table}

Kinro's three-month placement cohort shows how concentrated those economics are
at the low end. 71.6\% generate less than \$200 in annual
gross commission and 83.9\% generate less than \$300. The median is
\$116.40. These figures aggregate placed policies for each customer
and measure gross carrier commission before any producer payout.

The basic-service and high-touch broker thresholds in
Figure~\ref{fig:current-customer-commission-distribution} represent an
illustrative 50\% contribution-margin target before fixed overhead. Each
requires revenue equal to twice modeled direct cost, including the assumed
\$150 acquisition cost. The AI-enabled thresholds are direct-cost break-even:
five-year model usage plus \$150 in acquisition cost, divided across five
years. That is approximately \$41 annually with GPT-5.6 Sol and \$31 with
GPT-5.6 Luna. None of the four lines is an observed broker acceptance cutoff,
and the human-service target is a scenario rather than an estimate of a
universal required margin.

Within customers below \$100, the distribution is concentrated
near the top of the range: 74.4\% generate \$70--\textless\$100,
and the subgroup median is \$77.85.

The commission chart intentionally shows only the distribution of revenue.
The AI input- and output-token assumptions used to compare service models
remain in Appendix A.

\begin{figure}[H]
\centering
\begin{minipage}[c]{0.39\linewidth}
\centering
\begin{tikzpicture}[x=1cm,y=1cm]
  \path[fill=KinroLossOne,draw=white,line width=0.8pt] (0,0) -- (90:1.80) arc[start angle=90,end angle=-36.625,radius=1.80] -- cycle;
  \path[fill=KinroLossTwo,draw=white,line width=0.8pt] (0,0) -- (-36.625:1.80) arc[start angle=-36.625,end angle=-133.722,radius=1.80] -- cycle;
  \path[fill=KinroLossThree,draw=white,line width=0.8pt] (0,0) -- (-133.722:1.80) arc[start angle=-133.722,end angle=-183.123,radius=1.80] -- cycle;
  \path[fill=KinroLossFour,draw=white,line width=0.8pt] (0,0) -- (-183.123:1.80) arc[start angle=-183.123,end angle=-203.565,radius=1.80] -- cycle;
  \path[fill=KinroLossFive,draw=white,line width=0.8pt] (0,0) -- (-203.565:1.80) arc[start angle=-203.565,end angle=-219.464,radius=1.80] -- cycle;
  \path[fill=KinroLossSix,draw=white,line width=0.8pt] (0,0) -- (-219.464:1.80) arc[start angle=-219.464,end angle=-270,radius=1.80] -- cycle;

  \foreach \sliceangle/\sliceid in {26.688/1,-85.174/2,-158.423/3,-193.344/4,-211.515/5,-244.732/6}
    \node[font=\sffamily\bfseries\fontsize{7.2}{8}\selectfont,text=white] at (\sliceangle:1.25) {\sliceid};
\end{tikzpicture}
\end{minipage}\hfill
\begin{minipage}[c]{0.58\linewidth}
\raggedright\sffamily\fontsize{7.5}{9.2}\selectfont
\newcommand{\losslegend}[4]{%
  \tikz[baseline=-0.6ex]{\filldraw[fill=#1,draw=#1] (0,0) rectangle (0.18,0.18);}%
  \hspace{0.35em}\textbf{#2}\hspace{0.35em}#3\hfill\textbf{#4}\par\vspace{0.18em}}
\losslegend{KinroLossOne}{1}{Timing / already placed elsewhere}{35.17\%}
\losslegend{KinroLossTwo}{2}{Carrier appetite / coverage / product fit}{26.97\%}
\losslegend{KinroLossThree}{3}{Quote price competitiveness}{13.72\%}
\losslegend{KinroLossFour}{4}{Affordability / budget mismatch}{5.68\%}
\losslegend{KinroLossFive}{5}{Incumbent / broker preference}{4.42\%}
\losslegend{KinroLossSix}{6}{Other documented outcomes}{14.04\%}
\end{minipage}
\caption{\textbf{Kinro operational evidence.} Documented outcomes among current lost purchased leads. ``Timing / already placed elsewhere'' includes purchases elsewhere plus explicit slowness or missed deadlines; it signals urgency but does not establish causal delay. Snapshot: September 14, 2026.}
\label{fig:partner-lost-reasons}
\end{figure}
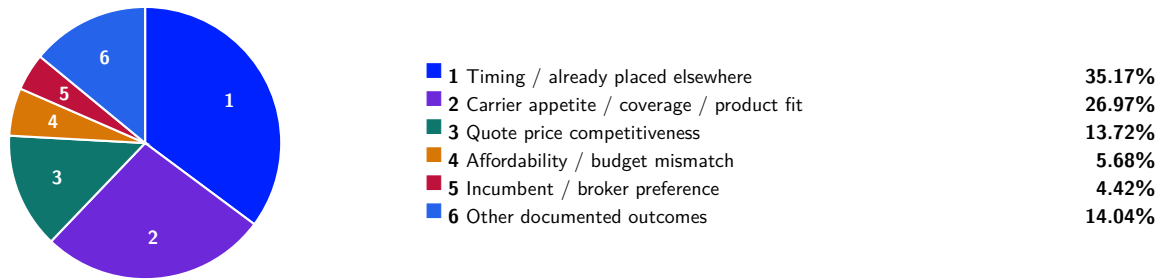

Figure~\ref{fig:partner-lost-reasons} also shows why speed matters. Among lost
leads with a specific documented outcome, 35.2\% had already placed coverage
elsewhere or explicitly cited slowness. Only 3.2\% explicitly documented
lateness or a missed deadline, so the figure does not establish that delay
caused the remaining losses. It does show that the opportunity to help an
owner can close quickly, making reliable follow-up and timely market search
part of the required service.

\Needspace{0.50\textheight}
\section{What an exceptional broker owes the
buyer}\label{what-an-exceptional-broker-owes-the-buyer}

An exceptional broker removes insurance from the owner's workload without
taking away control. The broker understands the business, gives buyer-side
advice, completes authorized work, and remains accountable for the result.

Although ``agent'' and ``broker'' are used loosely and legal roles can vary by
transaction, the principle here is simple: advice should begin with the
buyer's needs and compare suitable options across the markets the brokerage
can access. Sometimes the right recommendation is to keep the existing
policy, buy less coverage, consult a specialist, or not complete a sale.

That responsibility spans the entire relationship. Before the sale, the
broker must understand the business, explain the choices, search accessible
markets, and verify placement. After the sale, the broker must retain context,
service the policy, help with claims, and reassess coverage as the business
changes. Table~\ref{tab:five-year-cost} models the cost of providing that
attention at each stage.

Exceptional service means receiving the right help at the right moment,
engaging on one's own terms, never having to repeat context, remaining in
control of consequential decisions, and seeing each issue through to
completion. Its standard is the quality and continuity of customer
protection, not merely policies sold or tasks processed.

No person can remain continuously available, remember every unfinished
thread, and follow up reliably across an unlimited number of accounts. Human
experts therefore teach, evaluate, and govern the system while resolving
bounded exceptions. Universal service becomes economically possible as
demonstrated routine capabilities cease to require transaction-level human
review.

\clearpage
\section{How AI changes small-account
economics}\label{why-ai-changes-small-account-economics}

Consider a lawn-care business paying \$600 in annual premium and generating
\$90 in annual commission, or \$450 over five policy years. Table~\ref{tab:five-year-cost}
estimates five-year direct service costs of \$960 for basic human service and
\$2,400 for high-touch service, compared with modeled usage costs of \$52.80
for GPT-5.6 Sol and \$2.82 for GPT-5.6 Luna. The sales estimates allocate
full-funnel work, including work for prospects who do not ultimately purchase,
across each customer who binds. These are direct-cost comparisons,
not complete brokerage economics; Appendix A documents the assumptions and
exclusions.

\begin{figure}[H]
\centering
\resizebox{0.90\linewidth}{!}{\input{assets/human-touchpoint-budget.tex}}
\caption{\textbf{Illustrative model.} Five-year revenue from a \$90 annual-commission account: \$450. At a 50\% contribution-margin target, modeled acquisition and GPT-5.6 Sol usage costs leave room for only about two 15-minute human escalations. A high-touch broker costs \$2,400, or 5.3$\times$ this account's five-year revenue. Fixed overhead, governance, and licensing are excluded.}
\label{fig:human-touchpoint-budget}
\end{figure}
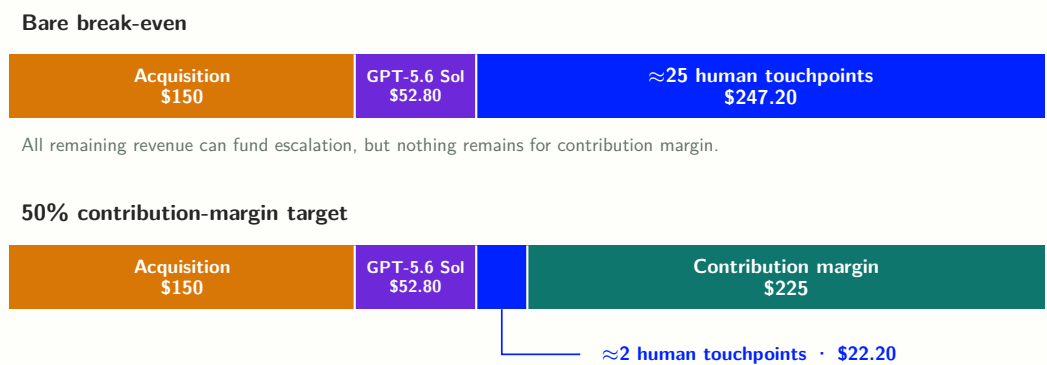

Low model cost creates room for continuous service, but every human escalation
consumes part of that advantage. In the conservative GPT-5.6 Sol case,
Figure~\ref{fig:human-touchpoint-budget} shows that a \$90 annual-commission
account can support only about two 15-minute human touchpoints over five years
while preserving a 50\% contribution margin.

AI-enabled service therefore becomes viable only when routine work can be
completed safely without transaction-level human review, while qualified
professionals remain available for consequential decisions and exceptions.
Lower cost alone does not establish service quality: broader authority should
depend on evidence that the brokerage applies insurance knowledge correctly,
respects customer permission, completes the work, and produces a verified
outcome.

Lower distribution cost cannot come at the expense of underwriting quality.
Even when carriers or MGAs retain authority over underwriting and pricing, an
AI-mediated sales process can affect loss performance through the accuracy of
risk classification, the completeness of submissions, and the mix of business
presented. Evaluation must therefore pair conversion, speed, cost, and customer
experience with submission quality, underwriting corrections and referrals,
and pricing adequacy.

Claims and loss ratios provide essential downstream evidence, but they mature
slowly and are influenced by pricing, geography, catastrophe exposure, and
business mix. Early evaluations should therefore measure leading indicators
such as material-fact completeness, classification accuracy, underwriter
corrections, and referral or decline rates. Claims frequency, severity, and
loss-ratio outcomes should be compared as credible cohorts mature.

\clearpage
\section{Conclusion and industry
agenda}\label{conclusion-and-industry-agenda}

Every small business should be able to obtain appropriate coverage and
competent continuing help. Yet the smallest policies rarely generate enough
commission to support the patience, education, continuity, and availability
owners need through human labor alone.

AI creates an opportunity to change those economics by sustaining
communication, memory, follow-through, and routine work across the customer
relationship. Human experts must continue to govern consequential decisions,
and the licensed brokerage must remain legally and financially accountable for
every outcome.

Realizing that opportunity safely requires brokers, carriers, MGAs, wholesale
markets, regulators, and consumer advocates to build a shared evidence base
through bounded real-world pilots, common benchmarks, and measurement of both
customer and underwriting outcomes.

\begin{center}
\small
\renewcommand{\arraystretch}{1.28}
\rowcolors{2}{white}{KinroPebble}
\begin{tabular}{@{}>{\raggedright\arraybackslash}p{0.24\linewidth} >{\raggedright\arraybackslash}p{0.70\linewidth}@{}}
\toprule
\textbf{Industry participant} & \textbf{How they can help} \\
\midrule
\textbf{Brokers} & Identify tasks AI could perform if proven sufficiently
accurate; contribute demonstrations, corrections, specialty knowledge, and
customer context; and define when qualified professionals must intervene. \\
\textbf{Carriers, MGAs, and wholesale markets} & Make appetite, coverage terms,
information requirements, and servicing rules understandable and accessible;
provide structured feedback on submission quality, underwriting decisions,
pricing, claims, and loss performance. \\
\textbf{Regulators} & Define requirements for disclosure, authorization,
accountability, monitoring, incident response, and remedies; create
evidence-based paths from bounded pilots to broader authority. Regulators are
already developing frameworks for AI governance, evaluation, and
accountability~\citep{naic-artificial-intelligence}. \\
\textbf{Consumer advocates} & Ensure benchmarks and deployment standards
reflect small-business owners' needs, including clarity, accessibility,
control, correction, and fair outcomes. \\
\bottomrule
\end{tabular}
\end{center}

Success would mean that every business receives exceptional insurance service.

\clearpage
\section{About Kinro}\label{about-kinro}

Kinro is a commercial insurance brokerage licensed in all 50 U.S. states and
offering most major lines of commercial insurance. Our goal is to provide the
best customer experience in the market to the long tail of businesses this
paper describes. Across the team, our insurance experience spans serving
micro-premium accounts and building insurance products and risk models from
the managing general agent (MGA) side of the industry. Our technical
experience includes building safety-critical systems for autonomous driving at
Zoox and training AI models at Google DeepMind to be more educational and
capable in finance.

\section*{Contributions and
Acknowledgments}\label{contributions-and-acknowledgments}
\addcontentsline{toc}{section}{Contributions and Acknowledgments}

\subsection*{Contributions}\label{contributions}
\addcontentsline{toc}{subsection}{Contributions}

The entire Kinro team contributed to this paper:

\begin{itemize}
\tightlist
\item
  Pierre-Alexandre Kamienny
\item
  Parthasarathi Ainampudi
\item
  Corentin Hugot
\item
  Hemanth Sai Kosari
\item
  Robert Martin
\item
  Armand Bechy
\end{itemize}

\subsection*{Acknowledgments}\label{acknowledgments}
\addcontentsline{toc}{subsection}{Acknowledgments}

We are grateful to Jonathan Crystal and Stephen McGovern of Crystal Venture
Partners, and to Bill Cecil, for generously sharing their time and expertise in
reviewing this paper.

\clearpage
\bibliographystyle{unsrtnat}
\bibliography{references}
\clearpage

\appendix
\renewcommand{\thesection}{Appendix \Alph{section}}

\section{Five-year human and AI workload assumptions}
\label{appendix-five-year-labor-model}

\subsection{Commission-distribution cohort and threshold construction}

Figure~\ref{fig:current-customer-commission-distribution} uses a production
snapshot as of September 13, 2026. Percentages use non-test customers with
policies Kinro placed during the previous three months and complete
commission-rate data. Gross commission is measured before producer payout.
The plotted AI-enabled thresholds sum five-year GPT-5.6 Sol or Luna usage and
the assumed \$150 acquisition cost, then divide by five years to show
direct-cost break-even. The basic-service and high-touch human thresholds
double the sum of five-year labor and acquisition cost, then divide by five
years to show a 50\% contribution-margin target. All four exclude fixed
overhead, infrastructure, and expert review; the human-service target is a
contribution margin, not a modeled net-profit margin.

The model follows one bound customer through five policy years while allocating
the work of the entire sales funnel across that customer. The five-year horizon
is an expected customer lifetime assumption, not an observed Kinro retention
result. Under a constant annual churn model, a five-year expected lifetime
implies 20\% annual churn and 80\% annual retention ($1 / 0.20 = 5$). At that
rate, 40.96\% of customers would be expected to reach the fifth policy year
($0.8^4$). The table uses the five-year expected lifetime as a single planning
case rather than probability-weighting each policy year. Its assumptions
combine an externally benchmarked labor rate, workload estimates grounded in
workflows Kinro has observed, and explicit service-design choices about cadence
and depth. They are planning estimates, not industry-wide averages or formal
time-and-motion measurements.

The \$40 rate is a rounded national, cross-state economic-cost proxy, not a
uniform wage. The U.S. Bureau of Labor Statistics reports median 2025 pay of
\$61,550 for insurance sales agents in agencies and brokerages, or about
\$29.60 per hour over 2,080 hours, including commissions and bonuses but excluding
self-employed owners~\citep{bls-insurance-sales-agents}. BLS also reports that
benefits were 30\% of private-industry employer compensation in June
2026~\citep{bls-employer-compensation}. Applying that broad benefit share implies
roughly \$42 per hour, so the model retains \$40 as a conservative assumption
rather than claiming false precision. Actual cost varies by state, role,
utilization, and compensation plan. Producer pay is often a share of agency
commission; the \$40 therefore values time and should not be added again where
the same producer compensation is already counted.

\begingroup\small\setlength{\tabcolsep}{4pt}\rowcolors{2}{white}{KinroPebble}
\begin{longtable}[]{@{}
  >{\raggedright\arraybackslash}p{0.22\linewidth}
  >{\raggedleft\arraybackslash}p{0.10\linewidth}
  >{\raggedleft\arraybackslash}p{0.10\linewidth}
  >{\raggedright\arraybackslash}p{0.50\linewidth}@{}}
\toprule\noalign{}
Component & Great & Minimal & Basis \\
\midrule\noalign{}
\endhead
\bottomrule\noalign{}
\endlastfoot
Sales and initial placement & 10h & 8h & Kinro full-funnel work estimate
allocated per bound customer. It includes conversations and follow-up with
prospects who do not purchase, plus education, market search, payment
confirmation, comparison, and verified bind for successful placements.
The basic-service broker spends less market-search time. \\
Four annual renewals & 8h & 4h & Kinro planning estimate: two hours per renewal
for the high-touch broker and one hour for the basic-service broker. \\
Certificates and routine service & 10h & 10h & Kinro planning estimate: two
hours per policy year for certificates, changes, payment follow-up, and routine
care. \\
Proactive customer check-ins & 10h & 0h & High-touch broker design choice: two
hours per policy year. The basic-service broker is reactive. \\
Proactive market checks & 20h & 0h & High-touch broker design choice: four one-hour
checks per year for five years. The basic-service broker does not proactively rescan. \\
Claims assistance & 2h & 2h & Scenario assumption: one claim in five years and
two hours to explain, report or coordinate, and follow up. \\
\midrule\noalign{}
\textbf{Total} & \textbf{60h (\$2,400)} & \textbf{24h (\$960)} & Labor
valued at \$40 per hour. Compliance is not included as a separate workload
component. \\
\end{longtable}
\endgroup

\subsection{Human-touchpoint affordability assumptions}

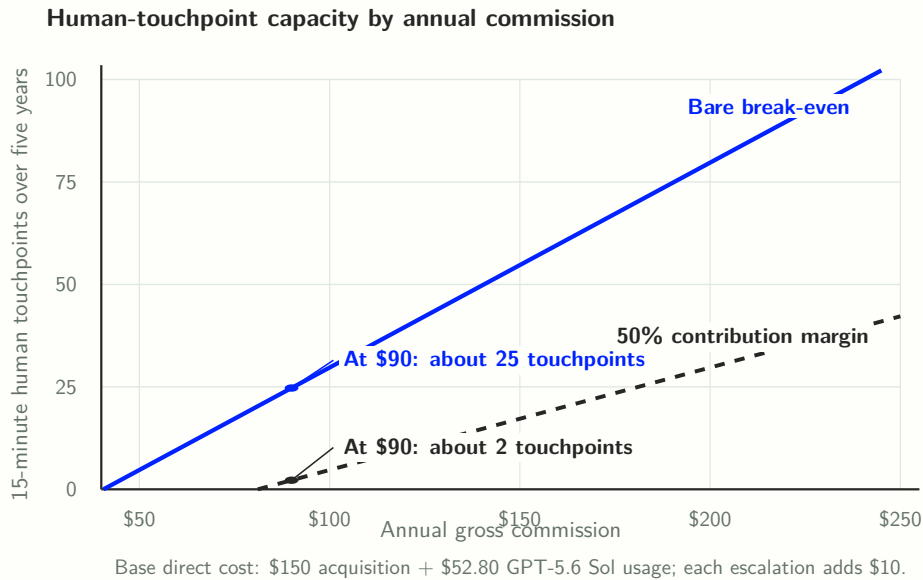
\begin{figure}[H]
\centering
\resizebox{0.78\linewidth}{!}{\input{assets/human-touchpoint-cost.tex}}
\caption{\textbf{Illustrative sensitivity analysis.} Maximum 15-minute human touchpoints over five years under bare break-even (solid) and a 50\% contribution-margin target (dashed).}
\label{fig:human-touchpoint-sensitivity}
\end{figure}

Figure~\ref{fig:human-touchpoint-budget} presents the \$90 annual-commission
case using the same five-year customer-lifecycle model as
Table~\ref{tab:five-year-cost}. Before
human escalation, the conservative GPT-5.6 Sol case has modeled AI-enabled
direct cost of \$202.80: \$150 for acquisition and \$52.80 for model usage.
Each human touchpoint is assigned 15 minutes and valued at \$10 using the same
\$40 hourly rate. For annual commission $c$, the bare break-even capacity is
$\max(0,(5c-202.80)/10)$ touchpoints. Reserving 50\% of revenue as contribution
margin limits modeled direct cost to the other half and reduces capacity to
$\max(0,(2.5c-202.80)/10)$. At \$90 in annual commission, the results are 24.72
and 2.22 touchpoints. Figure~\ref{fig:human-touchpoint-sensitivity} extends the
comparison across annual commission levels. This planning model excludes fixed
overhead, governance, and licensing; actual escalations occur in whole units
and may take more or less than 15 minutes.

\subsection{AI model-usage assumptions}

The \$52.80 Sol and \$2.82 Luna estimates are illustrative model usage, not
measured production averages or the full cost of AI-enabled service. GPT-5.6 Sol costs \$4 per million input tokens
and \$20 per million output tokens; GPT-5.6 Luna costs \$0.20 and \$1.20,
respectively~\citep{openai-sol-pricing,openai-luna-pricing}. Both cases retain the
main table's allowances:
2.0 million input and 200,000 output tokens for initial placement; 1.2 million
and 120,000 for renewals; 4.8 million and 480,000 for proactive market checks;
400,000 and 40,000 for customer check-ins; and 400,000 and 40,000 for
certificates, routine service, and one claim. The five-year total is 8.8
million input and 880,000 output tokens. These estimates exclude development,
infrastructure, carrier and vendor fees, licensing, governance, and human escalation.

\end{document}

%% file: assets/current-customer-annual-commission-distribution.tex
\begin{tikzpicture}[x=1cm,y=1cm]
  \begin{scope}[xshift=7.2cm]
  \draw[KinroPlotGrid,line width=0.35pt] (-0.10,0) -- (13.08,0);
  \node[font=\sffamily\fontsize{8.5}{9.4}\selectfont,anchor=east,text=KinroPlotLabel] at (-0.18,0) {0\%};
  \draw[KinroPlotGrid,line width=0.35pt] (-0.10,1) -- (13.08,1);
  \node[font=\sffamily\fontsize{8.5}{9.4}\selectfont,anchor=east,text=KinroPlotLabel] at (-0.18,1) {10\%};
  \draw[KinroPlotGrid,line width=0.35pt] (-0.10,2) -- (13.08,2);
  \node[font=\sffamily\fontsize{8.5}{9.4}\selectfont,anchor=east,text=KinroPlotLabel] at (-0.18,2) {20\%};
  \draw[KinroPlotGrid,line width=0.35pt] (-0.10,3) -- (13.08,3);
  \node[font=\sffamily\fontsize{8.5}{9.4}\selectfont,anchor=east,text=KinroPlotLabel] at (-0.18,3) {30\%};
  \draw[KinroPlotGrid,line width=0.35pt] (-0.10,4) -- (13.08,4);
  \node[font=\sffamily\fontsize{8.5}{9.4}\selectfont,anchor=east,text=KinroPlotLabel] at (-0.18,4) {40\%};
  \fill[KinroCobalt,rounded corners=0.5pt] (0.00,0) rectangle (0.78,3.79);
  \node[font=\sffamily\bfseries\fontsize{8.2}{9.0}\selectfont,anchor=south,text=KinroGraphite,fill=white,inner sep=0.8pt] at (0.39,3.85) {37.9\%};
  \node[font=\sffamily\fontsize{8.0}{8.8}\selectfont,anchor=north,text=KinroAsh] at (0.39,-0.10) {\textless\$100};
  \fill[KinroCobalt,rounded corners=0.5pt] (1.22,0) rectangle (2.00,3.37);
  \node[font=\sffamily\bfseries\fontsize{8.2}{9.0}\selectfont,anchor=south,text=KinroGraphite,fill=white,inner sep=0.8pt] at (1.61,3.43) {33.7\%};
  \node[font=\sffamily\fontsize{8.0}{8.8}\selectfont,anchor=north,text=KinroAsh] at (1.61,-0.10) {\$100s};
  \fill[KinroCobalt,rounded corners=0.5pt] (2.44,0) rectangle (3.22,1.23);
  \node[font=\sffamily\bfseries\fontsize{8.2}{9.0}\selectfont,anchor=south,text=KinroGraphite,fill=white,inner sep=0.8pt] at (2.83,1.29) {12.3\%};
  \node[font=\sffamily\fontsize{8.0}{8.8}\selectfont,anchor=north,text=KinroAsh] at (2.83,-0.10) {\$200s};
  \fill[KinroCobalt,rounded corners=0.5pt] (3.66,0) rectangle (4.44,0.39);
  \node[font=\sffamily\bfseries\fontsize{8.2}{9.0}\selectfont,anchor=south,text=KinroGraphite,fill=white,inner sep=0.8pt] at (4.05,0.45) {3.9\%};
  \node[font=\sffamily\fontsize{8.0}{8.8}\selectfont,anchor=north,text=KinroAsh] at (4.05,-0.10) {\$300s};
  \fill[KinroCobalt,rounded corners=0.5pt] (4.88,0) rectangle (5.66,0.36);
  \node[font=\sffamily\bfseries\fontsize{8.2}{9.0}\selectfont,anchor=south,text=KinroGraphite,fill=white,inner sep=0.8pt] at (5.27,0.42) {3.6\%};
  \node[font=\sffamily\fontsize{8.0}{8.8}\selectfont,anchor=north,text=KinroAsh] at (5.27,-0.10) {\$400s};
  \fill[KinroCobalt,rounded corners=0.5pt] (6.10,0) rectangle (6.88,0.16);
  \node[font=\sffamily\bfseries\fontsize{8.2}{9.0}\selectfont,anchor=south,text=KinroGraphite,fill=white,inner sep=0.8pt] at (6.49,0.22) {1.6\%};
  \node[font=\sffamily\fontsize{8.0}{8.8}\selectfont,anchor=north,text=KinroAsh] at (6.49,-0.10) {\$500s};
  \fill[KinroCobalt,rounded corners=0.5pt] (7.32,0) rectangle (8.10,0.03);
  \node[font=\sffamily\bfseries\fontsize{8.2}{9.0}\selectfont,anchor=south,text=KinroGraphite,fill=white,inner sep=0.8pt] at (7.71,0.09) {0.3\%};
  \node[font=\sffamily\fontsize{8.0}{8.8}\selectfont,anchor=north,text=KinroAsh] at (7.71,-0.10) {\$600s};
  \fill[KinroCobalt,rounded corners=0.5pt] (8.54,0) rectangle (9.32,0.06);
  \node[font=\sffamily\bfseries\fontsize{8.2}{9.0}\selectfont,anchor=south,text=KinroGraphite,fill=white,inner sep=0.8pt] at (8.93,0.12) {0.6\%};
  \node[font=\sffamily\fontsize{8.0}{8.8}\selectfont,anchor=north,text=KinroAsh] at (8.93,-0.10) {\$700s};
  \fill[KinroCobalt,rounded corners=0.5pt] (9.76,0) rectangle (10.54,0.03);
  \node[font=\sffamily\bfseries\fontsize{8.2}{9.0}\selectfont,anchor=south,text=KinroGraphite,fill=white,inner sep=0.8pt] at (10.15,0.09) {0.3\%};
  \node[font=\sffamily\fontsize{8.0}{8.8}\selectfont,anchor=north,text=KinroAsh] at (10.15,-0.10) {\$800s};
  \fill[KinroCobalt,rounded corners=0.5pt] (10.98,0) rectangle (11.76,0.19);
  \node[font=\sffamily\bfseries\fontsize{8.2}{9.0}\selectfont,anchor=south,text=KinroGraphite,fill=white,inner sep=0.8pt] at (11.37,0.25) {1.9\%};
  \node[font=\sffamily\fontsize{8.0}{8.8}\selectfont,anchor=north,text=KinroAsh] at (11.37,-0.10) {\$900s};
  \fill[KinroCobalt,rounded corners=0.5pt] (12.20,0) rectangle (12.98,0.39);
  \node[font=\sffamily\bfseries\fontsize{8.2}{9.0}\selectfont,anchor=south,text=KinroGraphite,fill=white,inner sep=0.8pt] at (12.59,0.45) {3.9\%};
  \node[font=\sffamily\fontsize{8.0}{8.8}\selectfont,anchor=north,text=KinroAsh] at (12.59,-0.10) {\textgreater\$1k};
  \draw[KinroLossFive,densely dashed,line width=1.65pt] (0.24,0) -- (0.24,4.08);
  \draw[KinroLossTwo,densely dashed,line width=1.65pt] (0.32,0) -- (0.32,4.08);
  \draw[KinroPlotCalls,densely dashed,line width=1.65pt] (5.22,0) -- (5.22,4.08);
  \draw[KinroLossThree,densely dashed,line width=1.65pt] (12.22,0) -- (12.22,4.08);
  \draw[KinroGraphite,line width=0.75pt] (0.00,0) rectangle (0.78,3.79);
  \node[font=\sffamily\fontsize{10.5}{11.4}\selectfont,text=KinroPlotLabel] at (6.49,-0.82) {Annual gross commission per customer};
  \node[rotate=90,font=\sffamily\fontsize{10.5}{11.4}\selectfont,text=KinroPlotLabel] at (-1.25,2.0) {Percentage of customers};
  \end{scope}

  \node[font=\sffamily\bfseries\fontsize{10.5}{11.5}\selectfont,anchor=center,text=KinroGraphite] at (10.00,5.32) {Illustrative annual-commission thresholds by service model};
  \draw[KinroLossFive,densely dashed,line width=1.65pt] (0.00,4.78) -- (0.36,4.78);
  \node[font=\sffamily\bfseries\fontsize{7.2}{8.0}\selectfont,anchor=west,align=left,text=KinroGraphite] at (0.48,4.78) {GPT-5.6 Luna};
  \draw[KinroLossTwo,densely dashed,line width=1.65pt] (5.05,4.78) -- (5.41,4.78);
  \node[font=\sffamily\bfseries\fontsize{7.2}{8.0}\selectfont,anchor=west,align=left,text=KinroGraphite] at (5.53,4.78) {GPT-5.6 Sol};
  \draw[KinroPlotCalls,densely dashed,line width=1.65pt] (10.00,4.78) -- (10.36,4.78);
  \node[font=\sffamily\bfseries\fontsize{7.2}{8.0}\selectfont,anchor=west,align=left,text=KinroGraphite] at (10.48,4.78) {Basic human service};
  \draw[KinroLossThree,densely dashed,line width=1.65pt] (14.75,4.78) -- (15.11,4.78);
  \node[font=\sffamily\bfseries\fontsize{7.2}{8.0}\selectfont,anchor=west,align=left,text=KinroGraphite] at (15.23,4.78) {High-touch human service};

  \filldraw[fill=white,draw=KinroRock,line width=0.65pt,rounded corners=0.8pt] (-0.25,0.50) rectangle (5.65,4.15);
  \node[font=\sffamily\bfseries\fontsize{8.2}{9.1}\selectfont,anchor=west,text=KinroGraphite] at (0.28,3.81) {Under \$100 \textperiodcentered\ \$10 bands};
  \node[font=\sffamily\fontsize{6.2}{7.0}\selectfont,anchor=west,text=KinroPlotLabel] at (0.28,3.55) {Share within the under-\$100 group};
  \foreach \yy/\lab in {1.25/0\%,1.92/10\%,2.59/20\%,3.26/30\%}{
    \draw[KinroPlotGrid,line width=0.3pt] (0.37,\yy) -- (5.17,\yy);
    \node[font=\sffamily\fontsize{5.8}{6.5}\selectfont,anchor=east,text=KinroPlotLabel] at (0.29,\yy) {\lab};
  }
  \foreach \xx/\hh/\pct/\band in {
    0.43/0.000/0/\$0,
    0.90/0.114/1.7/\$10,
    1.37/0.228/3.4/\$20,
    1.84/0.060/0.9/\$30,
    2.31/0.516/7.7/\$40,
    2.78/0.288/4.3/\$50,
    3.25/0.516/7.7/\$60,
    3.72/1.950/29.1/\$70,
    4.19/1.374/20.5/\$80,
    4.66/1.662/24.8/\$90}{
      \fill[KinroPlotSMS,rounded corners=0.3pt] (\xx,1.25) rectangle ({\xx+0.32},{1.25+\hh});
      \node[font=\sffamily\bfseries\fontsize{6.0}{6.8}\selectfont,anchor=south,text=KinroGraphite,fill=white,inner sep=0.35pt] at ({\xx+0.16},{1.28+\hh}) {\pct\%};
      \node[font=\sffamily\fontsize{6.0}{6.8}\selectfont,anchor=north,text=KinroAsh] at ({\xx+0.16},1.14) {\band};
  }
  \draw[KinroLossFive,densely dashed,line width=1.45pt] (1.86,1.25) -- (1.86,3.26);
  \draw[KinroLossTwo,densely dashed,line width=1.45pt] (2.34,1.25) -- (2.34,3.26);
  \draw[KinroRock,line width=0.45pt] (5.65,3.90) -- (7.20,3.79);
  \draw[KinroRock,line width=0.45pt] (5.65,0.65) -- (7.20,0.05);
\end{tikzpicture}

%% file: assets/self-employed-language-profile.tex
\begin{tikzpicture}[x=0.118cm,y=0.82cm]
  \tikzset{
    chartlabel/.style={font=\sffamily\fontsize{9}{10}\selectfont, text=KinroGraphite},
    barlabel/.style={font=\sffamily\bfseries\fontsize{8.5}{9.5}\selectfont, text=white},
    callout/.style={font=\sffamily\fontsize{8}{9}\selectfont, text=KinroPlotLabel}
  }

  \node[chartlabel,anchor=west] at (0,2.25) {Language spoken at home};
  \fill[KinroGraphite] (0,1.55) rectangle (74.6,2.05);
  \fill[KinroCobalt] (74.6,1.55) rectangle (89.01,2.05);
  \fill[KinroLossTwo] (89.01,1.55) rectangle (90.34,2.05);
  \fill[KinroPlotLabel] (90.34,1.55) rectangle (100,2.05);
  \node[barlabel] at (37.3,1.80) {English 74.6\%};
  \node[barlabel] at (81.805,1.80) {Spanish 14.4\%};
  \node[barlabel] at (95.17,1.80) {Other 9.7\%};
  \draw[KinroLossTwo,line width=0.45pt] (89.675,2.05) -- (89.675,2.42) -- (92.5,2.42);
  \node[callout,anchor=west] at (92.9,2.42) {Chinese 1.3\%};

  \node[chartlabel,anchor=west] at (0,1.10) {Reported English proficiency};
  \fill[KinroGraphite] (0,0.40) rectangle (87.82,0.90);
  \fill[KinroLossFour] (87.82,0.40) rectangle (94.24,0.90);
  \fill[KinroLossFive] (94.24,0.40) rectangle (98.53,0.90);
  \fill[KinroLossTwo] (98.53,0.40) rectangle (100,0.90);
  \node[barlabel] at (43.91,0.65) {English at home or very well 87.8\%};
  \draw[KinroPlotLabel,line width=0.45pt]
    (87.82,0.31) -- (87.82,0.23) -- (100,0.23) -- (100,0.31);
  \node[callout,anchor=north east] at (100,0.12) {12.2\% speak English less than very well};

  \node[callout,anchor=west] at (0,-0.62) {\textcolor{KinroLossFour}{\rule{6pt}{6pt}}\hspace{3pt}Well 6.4\%};
  \node[callout,anchor=west] at (27,-0.62) {\textcolor{KinroLossFive}{\rule{6pt}{6pt}}\hspace{3pt}Not well 4.3\%};
  \node[callout,anchor=west] at (54,-0.62) {\textcolor{KinroLossTwo}{\rule{6pt}{6pt}}\hspace{3pt}No English 1.5\%};
\end{tikzpicture}

%% file: assets/inbound-leads-by-local-week.tex
\definecolor{KinroHeatLow}{HTML}{DBEAFE}
\definecolor{KinroHeatHigh}{HTML}{087F5B}
\definecolor{KinroHeatZeroBorder}{HTML}{D8E2DC}
\noindent\makebox[\linewidth][r]{%
\begin{tikzpicture}[x=\dimexpr\linewidth-1cm\relax,y=0.44cm]
  \node[left,font=\sffamily\fontsize{7}{8}\selectfont\bfseries,text=KinroPlotLabel] at (-0.008,7) {Mon};
  \filldraw[fill=KinroHeatHigh!5.333333!KinroHeatLow,draw=white,line width=0.2pt,rounded corners=0.8pt] (0.0015,6.61) rectangle (0.040167,7.39);
  \filldraw[fill=KinroHeatHigh!2.666667!KinroHeatLow,draw=white,line width=0.2pt,rounded corners=0.8pt] (0.043167,6.61) rectangle (0.081833,7.39);
  \filldraw[fill=KinroHeatHigh!2.666667!KinroHeatLow,draw=white,line width=0.2pt,rounded corners=0.8pt] (0.084833,6.61) rectangle (0.1235,7.39);
  \filldraw[fill=KinroHeatHigh!6.666667!KinroHeatLow,draw=white,line width=0.2pt,rounded corners=0.8pt] (0.1265,6.61) rectangle (0.165167,7.39);
  \filldraw[fill=KinroHeatHigh!10.666667!KinroHeatLow,draw=white,line width=0.2pt,rounded corners=0.8pt] (0.168167,6.61) rectangle (0.206833,7.39);
  \filldraw[fill=KinroHeatHigh!9.333333!KinroHeatLow,draw=white,line width=0.2pt,rounded corners=0.8pt] (0.209833,6.61) rectangle (0.2485,7.39);
  \filldraw[fill=KinroHeatHigh!16!KinroHeatLow,draw=white,line width=0.2pt,rounded corners=0.8pt] (0.2515,6.61) rectangle (0.290167,7.39);
  \filldraw[fill=KinroHeatHigh!22.666667!KinroHeatLow,draw=white,line width=0.2pt,rounded corners=0.8pt] (0.293167,6.61) rectangle (0.331833,7.39);
  \filldraw[fill=KinroHeatHigh!72!KinroHeatLow,draw=white,line width=0.2pt,rounded corners=0.8pt] (0.334833,6.61) rectangle (0.3735,7.39);
  \filldraw[fill=KinroHeatHigh!80!KinroHeatLow,draw=white,line width=0.2pt,rounded corners=0.8pt] (0.3765,6.61) rectangle (0.415167,7.39);
  \filldraw[fill=KinroHeatHigh!70.666667!KinroHeatLow,draw=white,line width=0.2pt,rounded corners=0.8pt] (0.418167,6.61) rectangle (0.456833,7.39);
  \filldraw[fill=KinroHeatHigh!74.666667!KinroHeatLow,draw=white,line width=0.2pt,rounded corners=0.8pt] (0.459833,6.61) rectangle (0.4985,7.39);
  \filldraw[fill=KinroHeatHigh!61.333333!KinroHeatLow,draw=white,line width=0.2pt,rounded corners=0.8pt] (0.5015,6.61) rectangle (0.540167,7.39);
  \filldraw[fill=KinroHeatHigh!68!KinroHeatLow,draw=white,line width=0.2pt,rounded corners=0.8pt] (0.543167,6.61) rectangle (0.581833,7.39);
  \filldraw[fill=KinroHeatHigh!60!KinroHeatLow,draw=white,line width=0.2pt,rounded corners=0.8pt] (0.584833,6.61) rectangle (0.6235,7.39);
  \filldraw[fill=KinroHeatHigh!65.333333!KinroHeatLow,draw=white,line width=0.2pt,rounded corners=0.8pt] (0.6265,6.61) rectangle (0.665167,7.39);
  \filldraw[fill=KinroHeatHigh!76!KinroHeatLow,draw=white,line width=0.2pt,rounded corners=0.8pt] (0.668167,6.61) rectangle (0.706833,7.39);
  \filldraw[fill=KinroHeatHigh!61.333333!KinroHeatLow,draw=white,line width=0.2pt,rounded corners=0.8pt] (0.709833,6.61) rectangle (0.7485,7.39);
  \filldraw[fill=KinroHeatHigh!49.333333!KinroHeatLow,draw=white,line width=0.2pt,rounded corners=0.8pt] (0.7515,6.61) rectangle (0.790167,7.39);
  \filldraw[fill=KinroHeatHigh!37.333333!KinroHeatLow,draw=white,line width=0.2pt,rounded corners=0.8pt] (0.793167,6.61) rectangle (0.831833,7.39);
  \filldraw[fill=KinroHeatHigh!42.666667!KinroHeatLow,draw=white,line width=0.2pt,rounded corners=0.8pt] (0.834833,6.61) rectangle (0.8735,7.39);
  \filldraw[fill=KinroHeatHigh!26.666667!KinroHeatLow,draw=white,line width=0.2pt,rounded corners=0.8pt] (0.8765,6.61) rectangle (0.915167,7.39);
  \filldraw[fill=KinroHeatHigh!6.666667!KinroHeatLow,draw=white,line width=0.2pt,rounded corners=0.8pt] (0.918167,6.61) rectangle (0.956833,7.39);
  \filldraw[fill=KinroHeatHigh!13.333333!KinroHeatLow,draw=white,line width=0.2pt,rounded corners=0.8pt] (0.959833,6.61) rectangle (0.9985,7.39);
  \node[left,font=\sffamily\fontsize{7}{8}\selectfont\bfseries,text=KinroPlotLabel] at (-0.008,6) {Tue};
  \filldraw[fill=KinroHeatHigh!2.666667!KinroHeatLow,draw=white,line width=0.2pt,rounded corners=0.8pt] (0.0015,5.61) rectangle (0.040167,6.39);
  \filldraw[fill=KinroHeatHigh!1.333333!KinroHeatLow,draw=white,line width=0.2pt,rounded corners=0.8pt] (0.043167,5.61) rectangle (0.081833,6.39);
  \filldraw[fill=KinroHeatHigh!2.666667!KinroHeatLow,draw=white,line width=0.2pt,rounded corners=0.8pt] (0.084833,5.61) rectangle (0.1235,6.39);
  \filldraw[fill=KinroHeatHigh!2.666667!KinroHeatLow,draw=white,line width=0.2pt,rounded corners=0.8pt] (0.1265,5.61) rectangle (0.165167,6.39);
  \filldraw[fill=KinroHeatHigh!2.666667!KinroHeatLow,draw=white,line width=0.2pt,rounded corners=0.8pt] (0.168167,5.61) rectangle (0.206833,6.39);
  \filldraw[fill=KinroHeatHigh!6.666667!KinroHeatLow,draw=white,line width=0.2pt,rounded corners=0.8pt] (0.209833,5.61) rectangle (0.2485,6.39);
  \filldraw[fill=KinroHeatHigh!5.333333!KinroHeatLow,draw=white,line width=0.2pt,rounded corners=0.8pt] (0.2515,5.61) rectangle (0.290167,6.39);
  \filldraw[fill=KinroHeatHigh!17.333333!KinroHeatLow,draw=white,line width=0.2pt,rounded corners=0.8pt] (0.293167,5.61) rectangle (0.331833,6.39);
  \filldraw[fill=KinroHeatHigh!64!KinroHeatLow,draw=white,line width=0.2pt,rounded corners=0.8pt] (0.334833,5.61) rectangle (0.3735,6.39);
  \filldraw[fill=KinroHeatHigh!90.666667!KinroHeatLow,draw=white,line width=0.2pt,rounded corners=0.8pt] (0.3765,5.61) rectangle (0.415167,6.39);
  \filldraw[fill=KinroHeatHigh!72!KinroHeatLow,draw=white,line width=0.2pt,rounded corners=0.8pt] (0.418167,5.61) rectangle (0.456833,6.39);
  \filldraw[fill=KinroHeatHigh!77.333333!KinroHeatLow,draw=white,line width=0.2pt,rounded corners=0.8pt] (0.459833,5.61) rectangle (0.4985,6.39);
  \filldraw[fill=KinroHeatHigh!69.333333!KinroHeatLow,draw=white,line width=0.2pt,rounded corners=0.8pt] (0.5015,5.61) rectangle (0.540167,6.39);
  \filldraw[fill=KinroHeatHigh!56!KinroHeatLow,draw=white,line width=0.2pt,rounded corners=0.8pt] (0.543167,5.61) rectangle (0.581833,6.39);
  \filldraw[fill=KinroHeatHigh!66.666667!KinroHeatLow,draw=white,line width=0.2pt,rounded corners=0.8pt] (0.584833,5.61) rectangle (0.6235,6.39);
  \filldraw[fill=KinroHeatHigh!93.333333!KinroHeatLow,draw=white,line width=0.2pt,rounded corners=0.8pt] (0.6265,5.61) rectangle (0.665167,6.39);
  \filldraw[fill=KinroHeatHigh!66.666667!KinroHeatLow,draw=white,line width=0.2pt,rounded corners=0.8pt] (0.668167,5.61) rectangle (0.706833,6.39);
  \filldraw[fill=KinroHeatHigh!73.333333!KinroHeatLow,draw=white,line width=0.2pt,rounded corners=0.8pt] (0.709833,5.61) rectangle (0.7485,6.39);
  \filldraw[fill=KinroHeatHigh!52!KinroHeatLow,draw=white,line width=0.2pt,rounded corners=0.8pt] (0.7515,5.61) rectangle (0.790167,6.39);
  \filldraw[fill=KinroHeatHigh!37.333333!KinroHeatLow,draw=white,line width=0.2pt,rounded corners=0.8pt] (0.793167,5.61) rectangle (0.831833,6.39);
  \filldraw[fill=KinroHeatHigh!29.333333!KinroHeatLow,draw=white,line width=0.2pt,rounded corners=0.8pt] (0.834833,5.61) rectangle (0.8735,6.39);
  \filldraw[fill=KinroHeatHigh!26.666667!KinroHeatLow,draw=white,line width=0.2pt,rounded corners=0.8pt] (0.8765,5.61) rectangle (0.915167,6.39);
  \filldraw[fill=KinroHeatHigh!13.333333!KinroHeatLow,draw=white,line width=0.2pt,rounded corners=0.8pt] (0.918167,5.61) rectangle (0.956833,6.39);
  \filldraw[fill=KinroHeatHigh!8!KinroHeatLow,draw=white,line width=0.2pt,rounded corners=0.8pt] (0.959833,5.61) rectangle (0.9985,6.39);
  \node[left,font=\sffamily\fontsize{7}{8}\selectfont\bfseries,text=KinroPlotLabel] at (-0.008,5) {Wed};
  \filldraw[fill=KinroHeatHigh!6.666667!KinroHeatLow,draw=white,line width=0.2pt,rounded corners=0.8pt] (0.0015,4.61) rectangle (0.040167,5.39);
  \filldraw[fill=KinroHeatHigh!2.666667!KinroHeatLow,draw=white,line width=0.2pt,rounded corners=0.8pt] (0.043167,4.61) rectangle (0.081833,5.39);
  \filldraw[fill=KinroHeatHigh!4!KinroHeatLow,draw=white,line width=0.2pt,rounded corners=0.8pt] (0.084833,4.61) rectangle (0.1235,5.39);
  \filldraw[fill=white,draw=KinroHeatZeroBorder,line width=0.35pt,rounded corners=0.8pt] (0.1265,4.61) rectangle (0.165167,5.39);
  \filldraw[fill=KinroHeatHigh!6.666667!KinroHeatLow,draw=white,line width=0.2pt,rounded corners=0.8pt] (0.168167,4.61) rectangle (0.206833,5.39);
  \filldraw[fill=KinroHeatHigh!6.666667!KinroHeatLow,draw=white,line width=0.2pt,rounded corners=0.8pt] (0.209833,4.61) rectangle (0.2485,5.39);
  \filldraw[fill=KinroHeatHigh!12!KinroHeatLow,draw=white,line width=0.2pt,rounded corners=0.8pt] (0.2515,4.61) rectangle (0.290167,5.39);
  \filldraw[fill=KinroHeatHigh!17.333333!KinroHeatLow,draw=white,line width=0.2pt,rounded corners=0.8pt] (0.293167,4.61) rectangle (0.331833,5.39);
  \filldraw[fill=KinroHeatHigh!69.333333!KinroHeatLow,draw=white,line width=0.2pt,rounded corners=0.8pt] (0.334833,4.61) rectangle (0.3735,5.39);
  \filldraw[fill=KinroHeatHigh!66.666667!KinroHeatLow,draw=white,line width=0.2pt,rounded corners=0.8pt] (0.3765,4.61) rectangle (0.415167,5.39);
  \filldraw[fill=KinroHeatHigh!100!KinroHeatLow,draw=white,line width=0.2pt,rounded corners=0.8pt] (0.418167,4.61) rectangle (0.456833,5.39);
  \filldraw[fill=KinroHeatHigh!98.666667!KinroHeatLow,draw=white,line width=0.2pt,rounded corners=0.8pt] (0.459833,4.61) rectangle (0.4985,5.39);
  \filldraw[fill=KinroHeatHigh!84!KinroHeatLow,draw=white,line width=0.2pt,rounded corners=0.8pt] (0.5015,4.61) rectangle (0.540167,5.39);
  \filldraw[fill=KinroHeatHigh!66.666667!KinroHeatLow,draw=white,line width=0.2pt,rounded corners=0.8pt] (0.543167,4.61) rectangle (0.581833,5.39);
  \filldraw[fill=KinroHeatHigh!74.666667!KinroHeatLow,draw=white,line width=0.2pt,rounded corners=0.8pt] (0.584833,4.61) rectangle (0.6235,5.39);
  \filldraw[fill=KinroHeatHigh!69.333333!KinroHeatLow,draw=white,line width=0.2pt,rounded corners=0.8pt] (0.6265,4.61) rectangle (0.665167,5.39);
  \filldraw[fill=KinroHeatHigh!61.333333!KinroHeatLow,draw=white,line width=0.2pt,rounded corners=0.8pt] (0.668167,4.61) rectangle (0.706833,5.39);
  \filldraw[fill=KinroHeatHigh!56!KinroHeatLow,draw=white,line width=0.2pt,rounded corners=0.8pt] (0.709833,4.61) rectangle (0.7485,5.39);
  \filldraw[fill=KinroHeatHigh!38.666667!KinroHeatLow,draw=white,line width=0.2pt,rounded corners=0.8pt] (0.7515,4.61) rectangle (0.790167,5.39);
  \filldraw[fill=KinroHeatHigh!36!KinroHeatLow,draw=white,line width=0.2pt,rounded corners=0.8pt] (0.793167,4.61) rectangle (0.831833,5.39);
  \filldraw[fill=KinroHeatHigh!38.666667!KinroHeatLow,draw=white,line width=0.2pt,rounded corners=0.8pt] (0.834833,4.61) rectangle (0.8735,5.39);
  \filldraw[fill=KinroHeatHigh!22.666667!KinroHeatLow,draw=white,line width=0.2pt,rounded corners=0.8pt] (0.8765,4.61) rectangle (0.915167,5.39);
  \filldraw[fill=KinroHeatHigh!12!KinroHeatLow,draw=white,line width=0.2pt,rounded corners=0.8pt] (0.918167,4.61) rectangle (0.956833,5.39);
  \filldraw[fill=KinroHeatHigh!2.666667!KinroHeatLow,draw=white,line width=0.2pt,rounded corners=0.8pt] (0.959833,4.61) rectangle (0.9985,5.39);
  \node[left,font=\sffamily\fontsize{7}{8}\selectfont\bfseries,text=KinroPlotLabel] at (-0.008,4) {Thu};
  \filldraw[fill=KinroHeatHigh!6.666667!KinroHeatLow,draw=white,line width=0.2pt,rounded corners=0.8pt] (0.0015,3.61) rectangle (0.040167,4.39);
  \filldraw[fill=KinroHeatHigh!2.666667!KinroHeatLow,draw=white,line width=0.2pt,rounded corners=0.8pt] (0.043167,3.61) rectangle (0.081833,4.39);
  \filldraw[fill=KinroHeatHigh!4!KinroHeatLow,draw=white,line width=0.2pt,rounded corners=0.8pt] (0.084833,3.61) rectangle (0.1235,4.39);
  \filldraw[fill=KinroHeatHigh!2.666667!KinroHeatLow,draw=white,line width=0.2pt,rounded corners=0.8pt] (0.1265,3.61) rectangle (0.165167,4.39);
  \filldraw[fill=KinroHeatHigh!1.333333!KinroHeatLow,draw=white,line width=0.2pt,rounded corners=0.8pt] (0.168167,3.61) rectangle (0.206833,4.39);
  \filldraw[fill=KinroHeatHigh!6.666667!KinroHeatLow,draw=white,line width=0.2pt,rounded corners=0.8pt] (0.209833,3.61) rectangle (0.2485,4.39);
  \filldraw[fill=KinroHeatHigh!6.666667!KinroHeatLow,draw=white,line width=0.2pt,rounded corners=0.8pt] (0.2515,3.61) rectangle (0.290167,4.39);
  \filldraw[fill=KinroHeatHigh!17.333333!KinroHeatLow,draw=white,line width=0.2pt,rounded corners=0.8pt] (0.293167,3.61) rectangle (0.331833,4.39);
  \filldraw[fill=KinroHeatHigh!65.333333!KinroHeatLow,draw=white,line width=0.2pt,rounded corners=0.8pt] (0.334833,3.61) rectangle (0.3735,4.39);
  \filldraw[fill=KinroHeatHigh!60!KinroHeatLow,draw=white,line width=0.2pt,rounded corners=0.8pt] (0.3765,3.61) rectangle (0.415167,4.39);
  \filldraw[fill=KinroHeatHigh!73.333333!KinroHeatLow,draw=white,line width=0.2pt,rounded corners=0.8pt] (0.418167,3.61) rectangle (0.456833,4.39);
  \filldraw[fill=KinroHeatHigh!92!KinroHeatLow,draw=white,line width=0.2pt,rounded corners=0.8pt] (0.459833,3.61) rectangle (0.4985,4.39);
  \filldraw[fill=KinroHeatHigh!69.333333!KinroHeatLow,draw=white,line width=0.2pt,rounded corners=0.8pt] (0.5015,3.61) rectangle (0.540167,4.39);
  \filldraw[fill=KinroHeatHigh!73.333333!KinroHeatLow,draw=white,line width=0.2pt,rounded corners=0.8pt] (0.543167,3.61) rectangle (0.581833,4.39);
  \filldraw[fill=KinroHeatHigh!64!KinroHeatLow,draw=white,line width=0.2pt,rounded corners=0.8pt] (0.584833,3.61) rectangle (0.6235,4.39);
  \filldraw[fill=KinroHeatHigh!65.333333!KinroHeatLow,draw=white,line width=0.2pt,rounded corners=0.8pt] (0.6265,3.61) rectangle (0.665167,4.39);
  \filldraw[fill=KinroHeatHigh!58.666667!KinroHeatLow,draw=white,line width=0.2pt,rounded corners=0.8pt] (0.668167,3.61) rectangle (0.706833,4.39);
  \filldraw[fill=KinroHeatHigh!57.333333!KinroHeatLow,draw=white,line width=0.2pt,rounded corners=0.8pt] (0.709833,3.61) rectangle (0.7485,4.39);
  \filldraw[fill=KinroHeatHigh!52!KinroHeatLow,draw=white,line width=0.2pt,rounded corners=0.8pt] (0.7515,3.61) rectangle (0.790167,4.39);
  \filldraw[fill=KinroHeatHigh!24!KinroHeatLow,draw=white,line width=0.2pt,rounded corners=0.8pt] (0.793167,3.61) rectangle (0.831833,4.39);
  \filldraw[fill=KinroHeatHigh!40!KinroHeatLow,draw=white,line width=0.2pt,rounded corners=0.8pt] (0.834833,3.61) rectangle (0.8735,4.39);
  \filldraw[fill=KinroHeatHigh!17.333333!KinroHeatLow,draw=white,line width=0.2pt,rounded corners=0.8pt] (0.8765,3.61) rectangle (0.915167,4.39);
  \filldraw[fill=KinroHeatHigh!16!KinroHeatLow,draw=white,line width=0.2pt,rounded corners=0.8pt] (0.918167,3.61) rectangle (0.956833,4.39);
  \filldraw[fill=KinroHeatHigh!12!KinroHeatLow,draw=white,line width=0.2pt,rounded corners=0.8pt] (0.959833,3.61) rectangle (0.9985,4.39);
  \node[left,font=\sffamily\fontsize{7}{8}\selectfont\bfseries,text=KinroPlotLabel] at (-0.008,3) {Fri};
  \filldraw[fill=KinroHeatHigh!2.666667!KinroHeatLow,draw=white,line width=0.2pt,rounded corners=0.8pt] (0.0015,2.61) rectangle (0.040167,3.39);
  \filldraw[fill=KinroHeatHigh!2.666667!KinroHeatLow,draw=white,line width=0.2pt,rounded corners=0.8pt] (0.043167,2.61) rectangle (0.081833,3.39);
  \filldraw[fill=KinroHeatHigh!5.333333!KinroHeatLow,draw=white,line width=0.2pt,rounded corners=0.8pt] (0.084833,2.61) rectangle (0.1235,3.39);
  \filldraw[fill=KinroHeatHigh!6.666667!KinroHeatLow,draw=white,line width=0.2pt,rounded corners=0.8pt] (0.1265,2.61) rectangle (0.165167,3.39);
  \filldraw[fill=KinroHeatHigh!4!KinroHeatLow,draw=white,line width=0.2pt,rounded corners=0.8pt] (0.168167,2.61) rectangle (0.206833,3.39);
  \filldraw[fill=KinroHeatHigh!6.666667!KinroHeatLow,draw=white,line width=0.2pt,rounded corners=0.8pt] (0.209833,2.61) rectangle (0.2485,3.39);
  \filldraw[fill=KinroHeatHigh!10.666667!KinroHeatLow,draw=white,line width=0.2pt,rounded corners=0.8pt] (0.2515,2.61) rectangle (0.290167,3.39);
  \filldraw[fill=KinroHeatHigh!6.666667!KinroHeatLow,draw=white,line width=0.2pt,rounded corners=0.8pt] (0.293167,2.61) rectangle (0.331833,3.39);
  \filldraw[fill=KinroHeatHigh!61.333333!KinroHeatLow,draw=white,line width=0.2pt,rounded corners=0.8pt] (0.334833,2.61) rectangle (0.3735,3.39);
  \filldraw[fill=KinroHeatHigh!68!KinroHeatLow,draw=white,line width=0.2pt,rounded corners=0.8pt] (0.3765,2.61) rectangle (0.415167,3.39);
  \filldraw[fill=KinroHeatHigh!77.333333!KinroHeatLow,draw=white,line width=0.2pt,rounded corners=0.8pt] (0.418167,2.61) rectangle (0.456833,3.39);
  \filldraw[fill=KinroHeatHigh!80!KinroHeatLow,draw=white,line width=0.2pt,rounded corners=0.8pt] (0.459833,2.61) rectangle (0.4985,3.39);
  \filldraw[fill=KinroHeatHigh!77.333333!KinroHeatLow,draw=white,line width=0.2pt,rounded corners=0.8pt] (0.5015,2.61) rectangle (0.540167,3.39);
  \filldraw[fill=KinroHeatHigh!52!KinroHeatLow,draw=white,line width=0.2pt,rounded corners=0.8pt] (0.543167,2.61) rectangle (0.581833,3.39);
  \filldraw[fill=KinroHeatHigh!69.333333!KinroHeatLow,draw=white,line width=0.2pt,rounded corners=0.8pt] (0.584833,2.61) rectangle (0.6235,3.39);
  \filldraw[fill=KinroHeatHigh!80!KinroHeatLow,draw=white,line width=0.2pt,rounded corners=0.8pt] (0.6265,2.61) rectangle (0.665167,3.39);
  \filldraw[fill=KinroHeatHigh!57.333333!KinroHeatLow,draw=white,line width=0.2pt,rounded corners=0.8pt] (0.668167,2.61) rectangle (0.706833,3.39);
  \filldraw[fill=KinroHeatHigh!70.666667!KinroHeatLow,draw=white,line width=0.2pt,rounded corners=0.8pt] (0.709833,2.61) rectangle (0.7485,3.39);
  \filldraw[fill=KinroHeatHigh!44!KinroHeatLow,draw=white,line width=0.2pt,rounded corners=0.8pt] (0.7515,2.61) rectangle (0.790167,3.39);
  \filldraw[fill=KinroHeatHigh!32!KinroHeatLow,draw=white,line width=0.2pt,rounded corners=0.8pt] (0.793167,2.61) rectangle (0.831833,3.39);
  \filldraw[fill=KinroHeatHigh!32!KinroHeatLow,draw=white,line width=0.2pt,rounded corners=0.8pt] (0.834833,2.61) rectangle (0.8735,3.39);
  \filldraw[fill=KinroHeatHigh!20!KinroHeatLow,draw=white,line width=0.2pt,rounded corners=0.8pt] (0.8765,2.61) rectangle (0.915167,3.39);
  \filldraw[fill=KinroHeatHigh!12!KinroHeatLow,draw=white,line width=0.2pt,rounded corners=0.8pt] (0.918167,2.61) rectangle (0.956833,3.39);
  \filldraw[fill=KinroHeatHigh!5.333333!KinroHeatLow,draw=white,line width=0.2pt,rounded corners=0.8pt] (0.959833,2.61) rectangle (0.9985,3.39);
  \node[left,font=\sffamily\fontsize{7}{8}\selectfont\bfseries,text=KinroPlotLabel] at (-0.008,2) {Sat};
  \filldraw[fill=KinroHeatHigh!5.333333!KinroHeatLow,draw=white,line width=0.2pt,rounded corners=0.8pt] (0.0015,1.6099999999999999) rectangle (0.040167,2.39);
  \filldraw[fill=KinroHeatHigh!1.333333!KinroHeatLow,draw=white,line width=0.2pt,rounded corners=0.8pt] (0.043167,1.6099999999999999) rectangle (0.081833,2.39);
  \filldraw[fill=KinroHeatHigh!2.666667!KinroHeatLow,draw=white,line width=0.2pt,rounded corners=0.8pt] (0.084833,1.6099999999999999) rectangle (0.1235,2.39);
  \filldraw[fill=KinroHeatHigh!4!KinroHeatLow,draw=white,line width=0.2pt,rounded corners=0.8pt] (0.1265,1.6099999999999999) rectangle (0.165167,2.39);
  \filldraw[fill=KinroHeatHigh!1.333333!KinroHeatLow,draw=white,line width=0.2pt,rounded corners=0.8pt] (0.168167,1.6099999999999999) rectangle (0.206833,2.39);
  \filldraw[fill=white,draw=KinroHeatZeroBorder,line width=0.35pt,rounded corners=0.8pt] (0.209833,1.6099999999999999) rectangle (0.2485,2.39);
  \filldraw[fill=KinroHeatHigh!4!KinroHeatLow,draw=white,line width=0.2pt,rounded corners=0.8pt] (0.2515,1.6099999999999999) rectangle (0.290167,2.39);
  \filldraw[fill=KinroHeatHigh!9.333333!KinroHeatLow,draw=white,line width=0.2pt,rounded corners=0.8pt] (0.293167,1.6099999999999999) rectangle (0.331833,2.39);
  \filldraw[fill=KinroHeatHigh!52!KinroHeatLow,draw=white,line width=0.2pt,rounded corners=0.8pt] (0.334833,1.6099999999999999) rectangle (0.3735,2.39);
  \filldraw[fill=KinroHeatHigh!49.333333!KinroHeatLow,draw=white,line width=0.2pt,rounded corners=0.8pt] (0.3765,1.6099999999999999) rectangle (0.415167,2.39);
  \filldraw[fill=KinroHeatHigh!41.333333!KinroHeatLow,draw=white,line width=0.2pt,rounded corners=0.8pt] (0.418167,1.6099999999999999) rectangle (0.456833,2.39);
  \filldraw[fill=KinroHeatHigh!72!KinroHeatLow,draw=white,line width=0.2pt,rounded corners=0.8pt] (0.459833,1.6099999999999999) rectangle (0.4985,2.39);
  \filldraw[fill=KinroHeatHigh!57.333333!KinroHeatLow,draw=white,line width=0.2pt,rounded corners=0.8pt] (0.5015,1.6099999999999999) rectangle (0.540167,2.39);
  \filldraw[fill=KinroHeatHigh!30.666667!KinroHeatLow,draw=white,line width=0.2pt,rounded corners=0.8pt] (0.543167,1.6099999999999999) rectangle (0.581833,2.39);
  \filldraw[fill=KinroHeatHigh!20!KinroHeatLow,draw=white,line width=0.2pt,rounded corners=0.8pt] (0.584833,1.6099999999999999) rectangle (0.6235,2.39);
  \filldraw[fill=KinroHeatHigh!34.666667!KinroHeatLow,draw=white,line width=0.2pt,rounded corners=0.8pt] (0.6265,1.6099999999999999) rectangle (0.665167,2.39);
  \filldraw[fill=KinroHeatHigh!36!KinroHeatLow,draw=white,line width=0.2pt,rounded corners=0.8pt] (0.668167,1.6099999999999999) rectangle (0.706833,2.39);
  \filldraw[fill=KinroHeatHigh!46.666667!KinroHeatLow,draw=white,line width=0.2pt,rounded corners=0.8pt] (0.709833,1.6099999999999999) rectangle (0.7485,2.39);
  \filldraw[fill=KinroHeatHigh!29.333333!KinroHeatLow,draw=white,line width=0.2pt,rounded corners=0.8pt] (0.7515,1.6099999999999999) rectangle (0.790167,2.39);
  \filldraw[fill=KinroHeatHigh!32!KinroHeatLow,draw=white,line width=0.2pt,rounded corners=0.8pt] (0.793167,1.6099999999999999) rectangle (0.831833,2.39);
  \filldraw[fill=KinroHeatHigh!26.666667!KinroHeatLow,draw=white,line width=0.2pt,rounded corners=0.8pt] (0.834833,1.6099999999999999) rectangle (0.8735,2.39);
  \filldraw[fill=KinroHeatHigh!8!KinroHeatLow,draw=white,line width=0.2pt,rounded corners=0.8pt] (0.8765,1.6099999999999999) rectangle (0.915167,2.39);
  \filldraw[fill=KinroHeatHigh!6.666667!KinroHeatLow,draw=white,line width=0.2pt,rounded corners=0.8pt] (0.918167,1.6099999999999999) rectangle (0.956833,2.39);
  \filldraw[fill=KinroHeatHigh!6.666667!KinroHeatLow,draw=white,line width=0.2pt,rounded corners=0.8pt] (0.959833,1.6099999999999999) rectangle (0.9985,2.39);
  \node[left,font=\sffamily\fontsize{7}{8}\selectfont\bfseries,text=KinroPlotLabel] at (-0.008,1) {Sun};
  \filldraw[fill=KinroHeatHigh!4!KinroHeatLow,draw=white,line width=0.2pt,rounded corners=0.8pt] (0.0015,0.61) rectangle (0.040167,1.3900000000000001);
  \filldraw[fill=KinroHeatHigh!5.333333!KinroHeatLow,draw=white,line width=0.2pt,rounded corners=0.8pt] (0.043167,0.61) rectangle (0.081833,1.3900000000000001);
  \filldraw[fill=white,draw=KinroHeatZeroBorder,line width=0.35pt,rounded corners=0.8pt] (0.084833,0.61) rectangle (0.1235,1.3900000000000001);
  \filldraw[fill=white,draw=KinroHeatZeroBorder,line width=0.35pt,rounded corners=0.8pt] (0.1265,0.61) rectangle (0.165167,1.3900000000000001);
  \filldraw[fill=KinroHeatHigh!1.333333!KinroHeatLow,draw=white,line width=0.2pt,rounded corners=0.8pt] (0.168167,0.61) rectangle (0.206833,1.3900000000000001);
  \filldraw[fill=KinroHeatHigh!2.666667!KinroHeatLow,draw=white,line width=0.2pt,rounded corners=0.8pt] (0.209833,0.61) rectangle (0.2485,1.3900000000000001);
  \filldraw[fill=KinroHeatHigh!12!KinroHeatLow,draw=white,line width=0.2pt,rounded corners=0.8pt] (0.2515,0.61) rectangle (0.290167,1.3900000000000001);
  \filldraw[fill=KinroHeatHigh!6.666667!KinroHeatLow,draw=white,line width=0.2pt,rounded corners=0.8pt] (0.293167,0.61) rectangle (0.331833,1.3900000000000001);
  \filldraw[fill=KinroHeatHigh!20!KinroHeatLow,draw=white,line width=0.2pt,rounded corners=0.8pt] (0.334833,0.61) rectangle (0.3735,1.3900000000000001);
  \filldraw[fill=KinroHeatHigh!37.333333!KinroHeatLow,draw=white,line width=0.2pt,rounded corners=0.8pt] (0.3765,0.61) rectangle (0.415167,1.3900000000000001);
  \filldraw[fill=KinroHeatHigh!34.666667!KinroHeatLow,draw=white,line width=0.2pt,rounded corners=0.8pt] (0.418167,0.61) rectangle (0.456833,1.3900000000000001);
  \filldraw[fill=KinroHeatHigh!52!KinroHeatLow,draw=white,line width=0.2pt,rounded corners=0.8pt] (0.459833,0.61) rectangle (0.4985,1.3900000000000001);
  \filldraw[fill=KinroHeatHigh!38.666667!KinroHeatLow,draw=white,line width=0.2pt,rounded corners=0.8pt] (0.5015,0.61) rectangle (0.540167,1.3900000000000001);
  \filldraw[fill=KinroHeatHigh!44!KinroHeatLow,draw=white,line width=0.2pt,rounded corners=0.8pt] (0.543167,0.61) rectangle (0.581833,1.3900000000000001);
  \filldraw[fill=KinroHeatHigh!53.333333!KinroHeatLow,draw=white,line width=0.2pt,rounded corners=0.8pt] (0.584833,0.61) rectangle (0.6235,1.3900000000000001);
  \filldraw[fill=KinroHeatHigh!49.333333!KinroHeatLow,draw=white,line width=0.2pt,rounded corners=0.8pt] (0.6265,0.61) rectangle (0.665167,1.3900000000000001);
  \filldraw[fill=KinroHeatHigh!29.333333!KinroHeatLow,draw=white,line width=0.2pt,rounded corners=0.8pt] (0.668167,0.61) rectangle (0.706833,1.3900000000000001);
  \filldraw[fill=KinroHeatHigh!30.666667!KinroHeatLow,draw=white,line width=0.2pt,rounded corners=0.8pt] (0.709833,0.61) rectangle (0.7485,1.3900000000000001);
  \filldraw[fill=KinroHeatHigh!28!KinroHeatLow,draw=white,line width=0.2pt,rounded corners=0.8pt] (0.7515,0.61) rectangle (0.790167,1.3900000000000001);
  \filldraw[fill=KinroHeatHigh!21.333333!KinroHeatLow,draw=white,line width=0.2pt,rounded corners=0.8pt] (0.793167,0.61) rectangle (0.831833,1.3900000000000001);
  \filldraw[fill=KinroHeatHigh!25.333333!KinroHeatLow,draw=white,line width=0.2pt,rounded corners=0.8pt] (0.834833,0.61) rectangle (0.8735,1.3900000000000001);
  \filldraw[fill=KinroHeatHigh!13.333333!KinroHeatLow,draw=white,line width=0.2pt,rounded corners=0.8pt] (0.8765,0.61) rectangle (0.915167,1.3900000000000001);
  \filldraw[fill=KinroHeatHigh!6.666667!KinroHeatLow,draw=white,line width=0.2pt,rounded corners=0.8pt] (0.918167,0.61) rectangle (0.956833,1.3900000000000001);
  \filldraw[fill=KinroHeatHigh!6.666667!KinroHeatLow,draw=white,line width=0.2pt,rounded corners=0.8pt] (0.959833,0.61) rectangle (0.9985,1.3900000000000001);
  \node[below,font=\sffamily\fontsize{6.2}{7}\selectfont,text=KinroPlotLabel] at (0.020833,0.48) {12am};
  \node[below,font=\sffamily\fontsize{6.2}{7}\selectfont,text=KinroPlotLabel] at (0.104167,0.48) {2am};
  \node[below,font=\sffamily\fontsize{6.2}{7}\selectfont,text=KinroPlotLabel] at (0.1875,0.48) {4am};
  \node[below,font=\sffamily\fontsize{6.2}{7}\selectfont,text=KinroPlotLabel] at (0.270833,0.48) {6am};
  \node[below,font=\sffamily\fontsize{6.2}{7}\selectfont,text=KinroPlotLabel] at (0.354167,0.48) {8am};
  \node[below,font=\sffamily\fontsize{6.2}{7}\selectfont,text=KinroPlotLabel] at (0.4375,0.48) {10am};
  \node[below,font=\sffamily\fontsize{6.2}{7}\selectfont,text=KinroPlotLabel] at (0.520833,0.48) {12pm};
  \node[below,font=\sffamily\fontsize{6.2}{7}\selectfont,text=KinroPlotLabel] at (0.604167,0.48) {2pm};
  \node[below,font=\sffamily\fontsize{6.2}{7}\selectfont,text=KinroPlotLabel] at (0.6875,0.48) {4pm};
  \node[below,font=\sffamily\fontsize{6.2}{7}\selectfont,text=KinroPlotLabel] at (0.770833,0.48) {6pm};
  \node[below,font=\sffamily\fontsize{6.2}{7}\selectfont,text=KinroPlotLabel] at (0.854167,0.48) {8pm};
  \node[below,font=\sffamily\fontsize{6.2}{7}\selectfont,text=KinroPlotLabel] at (0.9375,0.48) {10pm};
  \node[anchor=east,font=\sffamily\fontsize{6.5}{7.5}\selectfont\bfseries,text=KinroPlotLabel] at (0.075,-0.48) {Relative activity};
  \path[fill=white,draw=none] (0.09,-0.66) rectangle (0.09635,-0.3);
  \path[fill=KinroHeatHigh!1.010101!KinroHeatLow,draw=none] (0.0962,-0.66) rectangle (0.10255,-0.3);
  \path[fill=KinroHeatHigh!2.020202!KinroHeatLow,draw=none] (0.1024,-0.66) rectangle (0.10875,-0.3);
  \path[fill=KinroHeatHigh!3.030303!KinroHeatLow,draw=none] (0.1086,-0.66) rectangle (0.11495,-0.3);
  \path[fill=KinroHeatHigh!4.040404!KinroHeatLow,draw=none] (0.1148,-0.66) rectangle (0.12115,-0.3);
  \path[fill=KinroHeatHigh!5.050505!KinroHeatLow,draw=none] (0.121,-0.66) rectangle (0.12735,-0.3);
  \path[fill=KinroHeatHigh!6.060606!KinroHeatLow,draw=none] (0.1272,-0.66) rectangle (0.13355,-0.3);
  \path[fill=KinroHeatHigh!7.070707!KinroHeatLow,draw=none] (0.1334,-0.66) rectangle (0.13975,-0.3);
  \path[fill=KinroHeatHigh!8.080808!KinroHeatLow,draw=none] (0.1396,-0.66) rectangle (0.14595,-0.3);
  \path[fill=KinroHeatHigh!9.090909!KinroHeatLow,draw=none] (0.1458,-0.66) rectangle (0.15215,-0.3);
  \path[fill=KinroHeatHigh!10.10101!KinroHeatLow,draw=none] (0.152,-0.66) rectangle (0.15835,-0.3);
  \path[fill=KinroHeatHigh!11.111111!KinroHeatLow,draw=none] (0.1582,-0.66) rectangle (0.16455,-0.3);
  \path[fill=KinroHeatHigh!12.121212!KinroHeatLow,draw=none] (0.1644,-0.66) rectangle (0.17075,-0.3);
  \path[fill=KinroHeatHigh!13.131313!KinroHeatLow,draw=none] (0.1706,-0.66) rectangle (0.17695,-0.3);
  \path[fill=KinroHeatHigh!14.141414!KinroHeatLow,draw=none] (0.1768,-0.66) rectangle (0.18315,-0.3);
  \path[fill=KinroHeatHigh!15.151515!KinroHeatLow,draw=none] (0.183,-0.66) rectangle (0.18935,-0.3);
  \path[fill=KinroHeatHigh!16.161616!KinroHeatLow,draw=none] (0.1892,-0.66) rectangle (0.19555,-0.3);
  \path[fill=KinroHeatHigh!17.171717!KinroHeatLow,draw=none] (0.1954,-0.66) rectangle (0.20175,-0.3);
  \path[fill=KinroHeatHigh!18.181818!KinroHeatLow,draw=none] (0.2016,-0.66) rectangle (0.20795,-0.3);
  \path[fill=KinroHeatHigh!19.191919!KinroHeatLow,draw=none] (0.2078,-0.66) rectangle (0.21415,-0.3);
  \path[fill=KinroHeatHigh!20.20202!KinroHeatLow,draw=none] (0.214,-0.66) rectangle (0.22035,-0.3);
  \path[fill=KinroHeatHigh!21.212121!KinroHeatLow,draw=none] (0.2202,-0.66) rectangle (0.22655,-0.3);
  \path[fill=KinroHeatHigh!22.222222!KinroHeatLow,draw=none] (0.2264,-0.66) rectangle (0.23275,-0.3);
  \path[fill=KinroHeatHigh!23.232323!KinroHeatLow,draw=none] (0.2326,-0.66) rectangle (0.23895,-0.3);
  \path[fill=KinroHeatHigh!24.242424!KinroHeatLow,draw=none] (0.2388,-0.66) rectangle (0.24515,-0.3);
  \path[fill=KinroHeatHigh!25.252525!KinroHeatLow,draw=none] (0.245,-0.66) rectangle (0.25135,-0.3);
  \path[fill=KinroHeatHigh!26.262626!KinroHeatLow,draw=none] (0.2512,-0.66) rectangle (0.25755,-0.3);
  \path[fill=KinroHeatHigh!27.272727!KinroHeatLow,draw=none] (0.2574,-0.66) rectangle (0.26375,-0.3);
  \path[fill=KinroHeatHigh!28.282828!KinroHeatLow,draw=none] (0.2636,-0.66) rectangle (0.26995,-0.3);
  \path[fill=KinroHeatHigh!29.292929!KinroHeatLow,draw=none] (0.2698,-0.66) rectangle (0.27615,-0.3);
  \path[fill=KinroHeatHigh!30.30303!KinroHeatLow,draw=none] (0.276,-0.66) rectangle (0.28235,-0.3);
  \path[fill=KinroHeatHigh!31.313131!KinroHeatLow,draw=none] (0.2822,-0.66) rectangle (0.28855,-0.3);
  \path[fill=KinroHeatHigh!32.323232!KinroHeatLow,draw=none] (0.2884,-0.66) rectangle (0.29475,-0.3);
  \path[fill=KinroHeatHigh!33.333333!KinroHeatLow,draw=none] (0.2946,-0.66) rectangle (0.30095,-0.3);
  \path[fill=KinroHeatHigh!34.343434!KinroHeatLow,draw=none] (0.3008,-0.66) rectangle (0.30715,-0.3);
  \path[fill=KinroHeatHigh!35.353535!KinroHeatLow,draw=none] (0.307,-0.66) rectangle (0.31335,-0.3);
  \path[fill=KinroHeatHigh!36.363636!KinroHeatLow,draw=none] (0.3132,-0.66) rectangle (0.31955,-0.3);
  \path[fill=KinroHeatHigh!37.373737!KinroHeatLow,draw=none] (0.3194,-0.66) rectangle (0.32575,-0.3);
  \path[fill=KinroHeatHigh!38.383838!KinroHeatLow,draw=none] (0.3256,-0.66) rectangle (0.33195,-0.3);
  \path[fill=KinroHeatHigh!39.393939!KinroHeatLow,draw=none] (0.3318,-0.66) rectangle (0.33815,-0.3);
  \path[fill=KinroHeatHigh!40.40404!KinroHeatLow,draw=none] (0.338,-0.66) rectangle (0.34435,-0.3);
  \path[fill=KinroHeatHigh!41.414141!KinroHeatLow,draw=none] (0.3442,-0.66) rectangle (0.35055,-0.3);
  \path[fill=KinroHeatHigh!42.424242!KinroHeatLow,draw=none] (0.3504,-0.66) rectangle (0.35675,-0.3);
  \path[fill=KinroHeatHigh!43.434343!KinroHeatLow,draw=none] (0.3566,-0.66) rectangle (0.36295,-0.3);
  \path[fill=KinroHeatHigh!44.444444!KinroHeatLow,draw=none] (0.3628,-0.66) rectangle (0.36915,-0.3);
  \path[fill=KinroHeatHigh!45.454545!KinroHeatLow,draw=none] (0.369,-0.66) rectangle (0.37535,-0.3);
  \path[fill=KinroHeatHigh!46.464646!KinroHeatLow,draw=none] (0.3752,-0.66) rectangle (0.38155,-0.3);
  \path[fill=KinroHeatHigh!47.474747!KinroHeatLow,draw=none] (0.3814,-0.66) rectangle (0.38775,-0.3);
  \path[fill=KinroHeatHigh!48.484848!KinroHeatLow,draw=none] (0.3876,-0.66) rectangle (0.39395,-0.3);
  \path[fill=KinroHeatHigh!49.494949!KinroHeatLow,draw=none] (0.3938,-0.66) rectangle (0.40015,-0.3);
  \path[fill=KinroHeatHigh!50.505051!KinroHeatLow,draw=none] (0.4,-0.66) rectangle (0.40635,-0.3);
  \path[fill=KinroHeatHigh!51.515152!KinroHeatLow,draw=none] (0.4062,-0.66) rectangle (0.41255,-0.3);
  \path[fill=KinroHeatHigh!52.525253!KinroHeatLow,draw=none] (0.4124,-0.66) rectangle (0.41875,-0.3);
  \path[fill=KinroHeatHigh!53.535354!KinroHeatLow,draw=none] (0.4186,-0.66) rectangle (0.42495,-0.3);
  \path[fill=KinroHeatHigh!54.545455!KinroHeatLow,draw=none] (0.4248,-0.66) rectangle (0.43115,-0.3);
  \path[fill=KinroHeatHigh!55.555556!KinroHeatLow,draw=none] (0.431,-0.66) rectangle (0.43735,-0.3);
  \path[fill=KinroHeatHigh!56.565657!KinroHeatLow,draw=none] (0.4372,-0.66) rectangle (0.44355,-0.3);
  \path[fill=KinroHeatHigh!57.575758!KinroHeatLow,draw=none] (0.4434,-0.66) rectangle (0.44975,-0.3);
  \path[fill=KinroHeatHigh!58.585859!KinroHeatLow,draw=none] (0.4496,-0.66) rectangle (0.45595,-0.3);
  \path[fill=KinroHeatHigh!59.59596!KinroHeatLow,draw=none] (0.4558,-0.66) rectangle (0.46215,-0.3);
  \path[fill=KinroHeatHigh!60.606061!KinroHeatLow,draw=none] (0.462,-0.66) rectangle (0.46835,-0.3);
  \path[fill=KinroHeatHigh!61.616162!KinroHeatLow,draw=none] (0.4682,-0.66) rectangle (0.47455,-0.3);
  \path[fill=KinroHeatHigh!62.626263!KinroHeatLow,draw=none] (0.4744,-0.66) rectangle (0.48075,-0.3);
  \path[fill=KinroHeatHigh!63.636364!KinroHeatLow,draw=none] (0.4806,-0.66) rectangle (0.48695,-0.3);
  \path[fill=KinroHeatHigh!64.646465!KinroHeatLow,draw=none] (0.4868,-0.66) rectangle (0.49315,-0.3);
  \path[fill=KinroHeatHigh!65.656566!KinroHeatLow,draw=none] (0.493,-0.66) rectangle (0.49935,-0.3);
  \path[fill=KinroHeatHigh!66.666667!KinroHeatLow,draw=none] (0.4992,-0.66) rectangle (0.50555,-0.3);
  \path[fill=KinroHeatHigh!67.676768!KinroHeatLow,draw=none] (0.5054,-0.66) rectangle (0.51175,-0.3);
  \path[fill=KinroHeatHigh!68.686869!KinroHeatLow,draw=none] (0.5116,-0.66) rectangle (0.51795,-0.3);
  \path[fill=KinroHeatHigh!69.69697!KinroHeatLow,draw=none] (0.5178,-0.66) rectangle (0.52415,-0.3);
  \path[fill=KinroHeatHigh!70.707071!KinroHeatLow,draw=none] (0.524,-0.66) rectangle (0.53035,-0.3);
  \path[fill=KinroHeatHigh!71.717172!KinroHeatLow,draw=none] (0.5302,-0.66) rectangle (0.53655,-0.3);
  \path[fill=KinroHeatHigh!72.727273!KinroHeatLow,draw=none] (0.5364,-0.66) rectangle (0.54275,-0.3);
  \path[fill=KinroHeatHigh!73.737374!KinroHeatLow,draw=none] (0.5426,-0.66) rectangle (0.54895,-0.3);
  \path[fill=KinroHeatHigh!74.747475!KinroHeatLow,draw=none] (0.5488,-0.66) rectangle (0.55515,-0.3);
  \path[fill=KinroHeatHigh!75.757576!KinroHeatLow,draw=none] (0.555,-0.66) rectangle (0.56135,-0.3);
  \path[fill=KinroHeatHigh!76.767677!KinroHeatLow,draw=none] (0.5612,-0.66) rectangle (0.56755,-0.3);
  \path[fill=KinroHeatHigh!77.777778!KinroHeatLow,draw=none] (0.5674,-0.66) rectangle (0.57375,-0.3);
  \path[fill=KinroHeatHigh!78.787879!KinroHeatLow,draw=none] (0.5736,-0.66) rectangle (0.57995,-0.3);
  \path[fill=KinroHeatHigh!79.79798!KinroHeatLow,draw=none] (0.5798,-0.66) rectangle (0.58615,-0.3);
  \path[fill=KinroHeatHigh!80.808081!KinroHeatLow,draw=none] (0.586,-0.66) rectangle (0.59235,-0.3);
  \path[fill=KinroHeatHigh!81.818182!KinroHeatLow,draw=none] (0.5922,-0.66) rectangle (0.59855,-0.3);
  \path[fill=KinroHeatHigh!82.828283!KinroHeatLow,draw=none] (0.5984,-0.66) rectangle (0.60475,-0.3);
  \path[fill=KinroHeatHigh!83.838384!KinroHeatLow,draw=none] (0.6046,-0.66) rectangle (0.61095,-0.3);
  \path[fill=KinroHeatHigh!84.848485!KinroHeatLow,draw=none] (0.6108,-0.66) rectangle (0.61715,-0.3);
  \path[fill=KinroHeatHigh!85.858586!KinroHeatLow,draw=none] (0.617,-0.66) rectangle (0.62335,-0.3);
  \path[fill=KinroHeatHigh!86.868687!KinroHeatLow,draw=none] (0.6232,-0.66) rectangle (0.62955,-0.3);
  \path[fill=KinroHeatHigh!87.878788!KinroHeatLow,draw=none] (0.6294,-0.66) rectangle (0.63575,-0.3);
  \path[fill=KinroHeatHigh!88.888889!KinroHeatLow,draw=none] (0.6356,-0.66) rectangle (0.64195,-0.3);
  \path[fill=KinroHeatHigh!89.89899!KinroHeatLow,draw=none] (0.6418,-0.66) rectangle (0.64815,-0.3);
  \path[fill=KinroHeatHigh!90.909091!KinroHeatLow,draw=none] (0.648,-0.66) rectangle (0.65435,-0.3);
  \path[fill=KinroHeatHigh!91.919192!KinroHeatLow,draw=none] (0.6542,-0.66) rectangle (0.66055,-0.3);
  \path[fill=KinroHeatHigh!92.929293!KinroHeatLow,draw=none] (0.6604,-0.66) rectangle (0.66675,-0.3);
  \path[fill=KinroHeatHigh!93.939394!KinroHeatLow,draw=none] (0.6666,-0.66) rectangle (0.67295,-0.3);
  \path[fill=KinroHeatHigh!94.949495!KinroHeatLow,draw=none] (0.6728,-0.66) rectangle (0.67915,-0.3);
  \path[fill=KinroHeatHigh!95.959596!KinroHeatLow,draw=none] (0.679,-0.66) rectangle (0.68535,-0.3);
  \path[fill=KinroHeatHigh!96.969697!KinroHeatLow,draw=none] (0.6852,-0.66) rectangle (0.69155,-0.3);
  \path[fill=KinroHeatHigh!97.979798!KinroHeatLow,draw=none] (0.6914,-0.66) rectangle (0.69775,-0.3);
  \path[fill=KinroHeatHigh!98.989899!KinroHeatLow,draw=none] (0.6976,-0.66) rectangle (0.70395,-0.3);
  \path[fill=KinroHeatHigh!100!KinroHeatLow,draw=none] (0.7038,-0.66) rectangle (0.71015,-0.3);
  \draw[KinroHeatZeroBorder,line width=0.4pt,rounded corners=0.8pt] (0.09,-0.66) rectangle (0.71,-0.3);
  \draw[KinroPlotLabel,line width=0.3pt] (0.09,-0.69) -- (0.09,-0.78);
  \node[below,font=\sffamily\fontsize{6}{6.8}\selectfont,text=KinroPlotLabel] at (0.09,-0.76) {0};
  \draw[KinroPlotLabel,line width=0.3pt] (0.152,-0.69) -- (0.152,-0.78);
  \node[below,font=\sffamily\fontsize{6}{6.8}\selectfont,text=KinroPlotLabel] at (0.152,-0.76) {0.1};
  \draw[KinroPlotLabel,line width=0.3pt] (0.214,-0.69) -- (0.214,-0.78);
  \node[below,font=\sffamily\fontsize{6}{6.8}\selectfont,text=KinroPlotLabel] at (0.214,-0.76) {0.2};
  \draw[KinroPlotLabel,line width=0.3pt] (0.276,-0.69) -- (0.276,-0.78);
  \node[below,font=\sffamily\fontsize{6}{6.8}\selectfont,text=KinroPlotLabel] at (0.276,-0.76) {0.3};
  \draw[KinroPlotLabel,line width=0.3pt] (0.338,-0.69) -- (0.338,-0.78);
  \node[below,font=\sffamily\fontsize{6}{6.8}\selectfont,text=KinroPlotLabel] at (0.338,-0.76) {0.4};
  \draw[KinroPlotLabel,line width=0.3pt] (0.4,-0.69) -- (0.4,-0.78);
  \node[below,font=\sffamily\fontsize{6}{6.8}\selectfont,text=KinroPlotLabel] at (0.4,-0.76) {0.5};
  \draw[KinroPlotLabel,line width=0.3pt] (0.462,-0.69) -- (0.462,-0.78);
  \node[below,font=\sffamily\fontsize{6}{6.8}\selectfont,text=KinroPlotLabel] at (0.462,-0.76) {0.6};
  \draw[KinroPlotLabel,line width=0.3pt] (0.524,-0.69) -- (0.524,-0.78);
  \node[below,font=\sffamily\fontsize{6}{6.8}\selectfont,text=KinroPlotLabel] at (0.524,-0.76) {0.7};
  \draw[KinroPlotLabel,line width=0.3pt] (0.586,-0.69) -- (0.586,-0.78);
  \node[below,font=\sffamily\fontsize{6}{6.8}\selectfont,text=KinroPlotLabel] at (0.586,-0.76) {0.8};
  \draw[KinroPlotLabel,line width=0.3pt] (0.648,-0.69) -- (0.648,-0.78);
  \node[below,font=\sffamily\fontsize{6}{6.8}\selectfont,text=KinroPlotLabel] at (0.648,-0.76) {0.9};
  \draw[KinroPlotLabel,line width=0.3pt] (0.71,-0.69) -- (0.71,-0.78);
  \node[below,font=\sffamily\fontsize{6}{6.8}\selectfont,text=KinroPlotLabel] at (0.71,-0.76) {1};
  \node[below,font=\sffamily\fontsize{6.2}{7}\selectfont,text=KinroPlotLabel] at (0.4,-1.3) {1 = busiest time};
\end{tikzpicture}%
}

%% file: assets/human-touchpoint-budget.tex
\begin{tikzpicture}[x=0.026cm,y=1cm]
  \fill[KinroOffWhite] (-18,0.95) rectangle (468,5.52);

  \node[anchor=west,font=\sffamily\bfseries\fontsize{7.2}{8}\selectfont,text=KinroGraphite]
    at (0,5.08) {Bare break-even};
  \fill[KinroLossFour] (0,4.0) rectangle (150,4.72);
  \fill[KinroLossTwo] (150,4.0) rectangle (202.8,4.72);
  \fill[KinroCobalt] (202.8,4.0) rectangle (450,4.72);
  \draw[white,line width=0.7pt] (150,4.0) -- (150,4.72);
  \draw[white,line width=0.7pt] (202.8,4.0) -- (202.8,4.72);
  \node[align=center,font=\sffamily\bfseries\fontsize{6.2}{6.9}\selectfont,text=white]
    at (75,4.36) {Acquisition\\\$150};
  \node[align=center,font=\sffamily\bfseries\fontsize{5.5}{6.1}\selectfont,text=white]
    at (176.4,4.36) {GPT-5.6 Sol\\\$52.80};
  \node[align=center,font=\sffamily\bfseries\fontsize{6.5}{7.2}\selectfont,text=white]
    at (326.4,4.36) {$\approx$25 human touchpoints\\\$247.20};
  \node[anchor=west,font=\sffamily\fontsize{6.1}{6.8}\selectfont,text=KinroPlotLabel]
    at (0,3.68) {All remaining revenue can fund escalation, but nothing remains for contribution margin.};

  \node[anchor=west,font=\sffamily\bfseries\fontsize{7.2}{8}\selectfont,text=KinroGraphite]
    at (0,2.92) {50\% contribution-margin target};
  \fill[KinroLossFour] (0,1.84) rectangle (150,2.56);
  \fill[KinroLossTwo] (150,1.84) rectangle (202.8,2.56);
  \fill[KinroCobalt] (202.8,1.84) rectangle (225,2.56);
  \fill[KinroLossThree] (225,1.84) rectangle (450,2.56);
  \draw[white,line width=0.7pt] (150,1.84) -- (150,2.56);
  \draw[white,line width=0.7pt] (202.8,1.84) -- (202.8,2.56);
  \draw[white,line width=0.7pt] (225,1.84) -- (225,2.56);
  \node[align=center,font=\sffamily\bfseries\fontsize{6.2}{6.9}\selectfont,text=white]
    at (75,2.20) {Acquisition\\\$150};
  \node[align=center,font=\sffamily\bfseries\fontsize{5.5}{6.1}\selectfont,text=white]
    at (176.4,2.20) {GPT-5.6 Sol\\\$52.80};
  \node[align=center,font=\sffamily\bfseries\fontsize{6.5}{7.2}\selectfont,text=white]
    at (337.5,2.20) {Contribution margin\\\$225};
  \draw[KinroCobalt,line width=0.55pt] (213.9,1.84) -- (213.9,1.33) -- (248,1.33);
  \node[anchor=west,font=\sffamily\bfseries\fontsize{6.2}{7}\selectfont,text=KinroCobalt]
    at (252,1.33) {$\approx$2 human touchpoints · \$22.20};

\end{tikzpicture}

%% file: assets/human-touchpoint-cost.tex
\begin{tikzpicture}[x=0.78cm,y=0.42cm]
  \fill[KinroOffWhite] (-1.35,-2.45) rectangle (11.35,12.1);

  \node[anchor=west,font=\sffamily\bfseries\fontsize{7.2}{8}\selectfont,text=KinroGraphite]
    at (-0.9,11.45) {Human-touchpoint capacity by annual commission};

  \foreach \y/\label in {0/0,2.5/25,5/50,7.5/75,10/100} {
    \draw[KinroPlotGrid,line width=0.4pt] (0,\y) -- (10.5,\y);
    \node[anchor=east,font=\sffamily\fontsize{6.2}{7}\selectfont,text=KinroPlotLabel]
      at (-0.15,\y) {\label};
  }
  \foreach \x/\label in {0.5/\$50,3/\$100,5.5/\$150,8/\$200,10.5/\$250} {
    \draw[KinroPlotGrid,line width=0.35pt] (\x,0) -- (\x,10);
    \node[anchor=north,font=\sffamily\fontsize{6.2}{7}\selectfont,text=KinroPlotLabel]
      at (\x,-0.18) {\label};
  }

  \draw[KinroGraphite,line width=0.65pt] (0,0) -- (10.75,0);
  \draw[KinroGraphite,line width=0.65pt] (0,0) -- (0,10.35);
  \node[font=\sffamily\fontsize{6.7}{7.5}\selectfont,text=KinroPlotLabel]
    at (5.25,-1.02) {Annual gross commission};
  \node[rotate=90,font=\sffamily\fontsize{6.7}{7.5}\selectfont,text=KinroPlotLabel]
    at (-1.05,5) {15-minute human touchpoints over five years};

  \draw[KinroCobalt,line width=1.25pt] (0.028,0) -- (10.25,10.222);
  \draw[KinroGraphite,dashed,line width=1.15pt] (2.056,0) -- (10.5,4.222);

  \node[anchor=east,fill=KinroOffWhite,inner sep=1.2pt,font=\sffamily\bfseries\fontsize{6.4}{7.2}\selectfont,text=KinroCobalt]
    at (9.9,9.35) {Bare break-even};
  \node[anchor=east,fill=KinroOffWhite,inner sep=1.2pt,font=\sffamily\bfseries\fontsize{6.4}{7.2}\selectfont,text=KinroGraphite]
    at (10.15,3.72) {50\% contribution margin};

  \fill[KinroCobalt] (2.5,2.472) circle (0.09);
  \draw[KinroCobalt,line width=0.45pt] (2.5,2.472) -- (3.05,3.15);
  \node[anchor=west,fill=KinroOffWhite,inner sep=1.2pt,font=\sffamily\bfseries\fontsize{6.4}{7.2}\selectfont,text=KinroCobalt]
    at (3.12,3.15) {At \$90: about 25 touchpoints};

  \fill[KinroGraphite] (2.5,0.222) circle (0.09);
  \draw[KinroGraphite,line width=0.45pt] (2.5,0.222) -- (3.05,1.02);
  \node[anchor=west,fill=KinroOffWhite,inner sep=1.2pt,font=\sffamily\bfseries\fontsize{6.4}{7.2}\selectfont,text=KinroGraphite]
    at (3.12,1.02) {At \$90: about 2 touchpoints};

  \node[anchor=west,font=\sffamily\fontsize{6.1}{7}\selectfont,text=KinroPlotLabel]
    at (0,-1.92) {Base direct cost: \$150 acquisition + \$52.80 GPT-5.6 Sol usage; each escalation adds \$10.};
\end{tikzpicture}